\documentclass{emulateapj}

\newcommand{\SR}{Strehl ratio}
\newcommand{\XAO}{extreme-AO}

\shorttitle{BIGRE}
\shortauthors{ANTICHI ET AL.}

\begin{document}

\title{BIGRE: a low cross-talk integral field unit tailored for extrasolar planets imaging spectroscopy}

\author{Jacopo Antichi}
\affil{LAOG-Laboratoire d'Astrophysique de Grenoble, B.P. 53, F-38041 Grenoble Cedex 9, France;\\
INAF-Osservatorio Astronomico di Padova, Vicolo dell'Osservatorio 5, I-35122 Padova, Italy}

\author{Kjetil Dohlen}
\affil{LAM-Laboratoire d'Astrophysique de Marseille, B.P. 8, F-13376 Marseille Cedex 12, France}

\author{Raffaele G. Gratton, Dino Mesa, Riccardo U. Claudi, Enrico Giro}
\affil{INAF-Osservatorio Astronomico di Padova, Vicolo dell'Osservatorio 5, I-35122 Padova, Italy}

\author{Anthony Boccaletti}
\affil{LESIA-Laboratoire d'Etudes Spatiales et d'Instrumentation en Astrophysique, F-92190 Meudon, France}

\author{David Mouillet, Pascal Puget, Jean-Luc Beuzit}
\affil{LAOG-Laboratoire d'Astrophysique de Grenoble, B.P. 53, F-38041 Grenoble Cedex 9, France}

\begin{abstract}
Integral field spectroscopy (IFS) represents a powerful technique for the detection and characterization
of extrasolar planets through high contrast imaging, since it allows to obtain simultaneously a large
number of monochromatic images. These can be used to calibrate and then to reduce the impact of speckles,
once their chromatic dependence is taken into account. The main concern in designing
integral field spectrographs for high contrast imaging is the impact of the diffraction effects and
the non-common path aberrations together with an efficient use of the detector pixels.
We focus our attention on integral field spectrographs based on lenslet-arrays, discussing the main
features of these designs: the conditions of appropriate spatial and spectral sampling of the resulting
spectrograph's slit functions and their related cross-talk terms when the system works at the
diffraction limit. We present a new scheme for the integral field unit (IFU) based on a dual-lenslet
device (BIGRE), that solves some of the problems related to the classical TIGER design when used for such applications.
We show that BIGRE provides much lower cross-talk signals than TIGER, allowing a more efficient use of the detector
pixels and a considerable saving of the overall cost of a lenslet-based integral field spectrograph.
\end{abstract}


\section{Introduction}\label{sec:Intro}

Imaging of a significant number of extrasolar planets requires achieving star vs. planet contrasts
of $\sim 10^6$\ (young giant planets), or even $10^8-10^{10}$\ (old giant and rocky planets)
at a few tenths of an arc-second from a star, which value is $\sim 10 \cdot \lambda/D$ in the near
infrared for telescopes having pupil sizes of $D$ $\sim 10$~m.

In this regime, the dominant noise
contribution is due to the stellar background. To achieve these ambitious goals, high contrast imagers
usually include various components. First, an extreme adaptive optics (XAO) system is used, allowing to
correct aberrations up to a high order, and providing a high Strehl Ratio (SR). Second, some
coronagraph is included, attenuating the coherent diffraction pattern of the on-axis point spread
function (PSF). Proper combination of these two devices allows reduction of the stellar
background down to values of $\sim 10^{-5}$\ out to the AO
system control radius\footnote{$\propto 1/2d$, $d \equiv$ actuator spacing projected on the telescope pupil.},
for state-of-the-art system. This background is due to a rapidly changing halo of speckles generated
by residual telescope pupil phase distortions, that have spacial frequencies close to those of planet
images. In order to avoid false alarms, the detection threshold level should then be set at several
times the root mean square (RMS) noise level.

Even in the favourable case where the speckle intensity
distribution can be assumed to be Gaussian \citep{M08a} the detection confidence limit should be at least 5 times the
noise level. This implies that at angular separations $\le 10 \cdot \lambda/D$, the limiting contrast
provided by state-of-the-art {\XAO} and coronagraphy is $\sim 2 \cdot 10^{4}$\ for 8-10 m telescopes.
In addition, phase aberrations originating inside the optical train not corrected by the
{\XAO} system produce speckles of longer lifetime (minutes or hours) than those due the atmosphere.
Other slowly varying (of the order of seconds) phase errors are due to aliasing effects in the
wavefront sensor \citep{PM04} and --- for coronagraphic systems --- to adaptive optics time-lag
\citep{MC05}.

Beyond a handful of favorable cases where planets are warm - e.g. \cite{CHA04, CHA05, N05} -
or with large separation from their parent star \citep{K08}, or eventually with both these properties
\citep{Laf08, M08b}, additional techniques are required to reach the larger contrasts needed for extrasolar planets detection.

Simultaneous differential imaging (SDI) is a high-contrast imaging differential technique by which subtraction
of different images of the same field acquired simultaneously by the same instrument allows to remove or reduce
the noise produced by atmospheric and instrumental phase aberrations. The SDI principle can be applied to images
obtained with different polarization modes \citep{GI04} or selecting two distinct wavelengths in a fixed spectral
range \citep{L05, M05}, or better exploiting the entire spectral range by integral field spectroscopy \citep{BAl06}.
In this paper we will focus on SDI based upon this latter strategy only.

Essentially, SDI is a calibration technique \citep{S87, R99, M00, SF02, BBe04, BAl06, RW06, T07}: images are acquired
simultaneously in bands at close wavelengths where the planetary (but not the stellar) flux differ appreciably.
Subtracting each other these images should allow to remove or at least reduce the speckle noise, since this is
assumed to be similar in the various images after a suitable chromatic re-scaling, while the planet signal is left
nearly untouched.

\begin{figure}[!ht]
\begin{center}
\resizebox{0.48\textwidth}{!}{\includegraphics{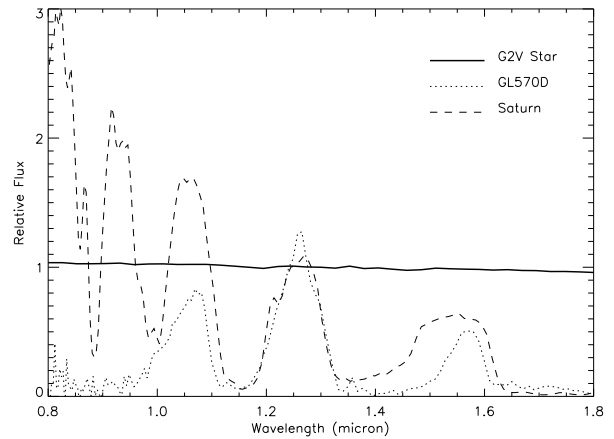}}
\caption{Near infrared low resolution spectra (two-pixel resolving power $R = 50$) of the Brown Dwarf GL 570 D, of the planet Saturn
and of a G2V star (by courtesy of Dr. L. Testi and Dr. F. Ghinassi) obtained at TNG+NICS with its Amici Prism.
The spectra are normalized to their flux at $1.30 \ \mu m$.}
\label{fg:Testi}
\end{center}
\end{figure}

There at least two ways to exploit this calibration technique.
In the more traditional approach, specific characteristics of the (expected) planetary spectra are exploited.
As indicated by various theoretical work \citep{S00, B02, BAd03, S03, BAd04} and observations
(e.g. of Brown Dwarfs and gaseous planets in the Solar System), the spectra of giant planets are dominated by several absorbtion bands (mainly due to methane and water vapor) at both visible and near infrared wavelengths.
In such a case, SDI may work by subtracting images where the planet signal
is absent from those where it is present, while the background is nearly the same, because the spectrum of the parent star is nearly
featureless, see Figure \ref{fg:Testi}. The main advantage of this technique, is the minimum assumptions required on
the chromatic behavior of speckles; however, this technique allows only a limited reduction of noise.
Alternatively, we might try to model the variation of speckles with wavelength \citep{SF02}.
In principle this allows to remove completely speckle noise without making any assumption about
the planetary spectrum, hence allowing to retrieve the real planetary spectrum \citep{T07}.

Independently to the adopted SDI recipe, integral field spectrograph designs tuned for diffraction-limited high-contrast
imaging should take into account several effects jeopardizing the interpolation procedures requested before simultaneous spectral
subtractions, which in turn severely limit the accuracy of this calibration technique.

In this paper we present a discussion of these effects and derive the basic equations that should be considered
when designing lenslet-based diffraction-limited integral field spectrographs. Then, we describe
a new concept for the lenslet-array shaping the IFU of such instruments (i.e. BIGRE)
allowing to improve significantly over the main limitations of the more traditional designs based
on the TIGER concept.

The structure of the paper is as follows. In \S \ \ref{sec:Speckle} we recall the basics of a
post-coronagraphic speckle field. In \S \ \ref{sec:SDI} we summarize the principle of SDI.
In \S \ \ref{sec:S-SDI} we discuss the basics of spectroscopic SDI (hereafter S-SDI),
defining the conditions allowing to avoid aliasing errors when sampling both the entrance speckle field
and the final exit slit functions. In \S \ \ref{sec:IFS-Options} we present various options for IFS-instruments
suited for S-SDI. In \S \ \ref{sec:CT} we define the cross-talk terms in the case of diffraction-limited lenslet-based IFS.
In \S \ \ref{sec:TIGER} we derive the rules governing the image propagation at the diffraction limit through
the TIGER concept, and in \S \ \ref{sec:BIGRE} the ones proper to the new BIGRE concept. Specifically,
we explain here how to conceive a BIGRE-oriented IFS instrument adopting standard dioptric devices.
In \S \ \ref{sec:SPHERE-IFS} we present two design setups (based on BIGRE and TIGER respectively)
for SPHERE\footnote{SPHERE is an instrument designed and built by a consortium of LAOG,
MPIA, LAM, LESIA, LUAN, INAF, Observatoire de Gen\`eve, ETH, NOVA, ONERA and ASTRON in
collaboration with and under from ESO. Its science objective is the direct detection and characterization
of giant extrasolar planets in the visible and near-infrared \citep{BJL08}.}, indicating the solution adopted for its
future IFS. In \S \ \ref{sec:TIGERvsBIGRE} we compare the TIGER and the BIGRE concepts in terms of coherent and incoherent
signals suppression, considering several cases for the single lens shape and the IFU lattice configuration. Finally,
our conclusions are drawn in \S \ \ref{sec:end}.

\section{Post-coronagraphic speckle field modeling}\label{sec:Speckle}

An appropriate understanding of chromatic intensity (e.g. \cite{R99}) and spatial (e.g. \cite{SF02}) scaling
of a speckle field is basic to any application of the SDI calibration technique.
For this reason a short description of these physical concepts is fundamental to introduce the reader to the topics treated in
the rest of the paper. Inspired by the approach of \cite{P03}, we will use the Fraunhofer approximation to
describe the impact of small residual phase variations of the electric field $(e)$ imaged
on a fixed post-coronagraphic entrance pupil plane\footnote{Hereafter Fourier pairs are defined with the same
letter written in small and capital case respectively.}, i.e. the working case of high-contrast imaging
instruments like SPHERE.

While this approach allows a simple mathematical treatment and physical
understanding, it ignores more complex effects due to amplitude errors and Fresnel propagation, as pointed out
by \cite{M06}. It is outside the scope of this paper to discuss such effects, which can be minimized by careful
instrument design, but it is likely that they will set the ultimate limit of planet imaging.

\begin{figure}[!ht]
\begin{center}
\resizebox{0.48\textwidth}{!}{\includegraphics{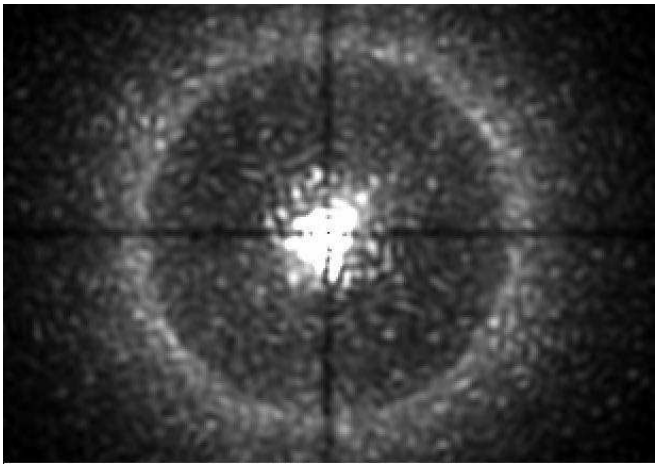}}
\caption{Example of a post-coronagraphic speckle field (integration time = $0.5 \ msec$ , wavelength = $1
\ \mu m$), simulating the {\XAO} system and the 4 quadrant phase mask coronagraph of SPHERE
(by courtesy of the SPHERE team).}
\label{fg:speckle-image}
\end{center}
\end{figure}

The most general expression of the monochromatic electric field once projected on the coronagraphic entrance pupil plane is:

\begin{equation}\label{eq:Coro-1}
e \equiv p \cdot exp[i \cdot \phi],
\end{equation}

\noindent where $(p)$ is the coronagraphic pupil transmission function, and $(\phi)$ is the phase of the electric field
evaluated over this coronagraphic pupil plane. Assuming a perfect optical propagation from the telescope to this plane
- i.e. no differential chromatical aberrations in the beam - the chromatism of the phase can be written explicitly as a function
of the wavelength $(\lambda)$ and the wavefront error $(w)$ as follows:

\begin{equation}\label{eq:Coro-2}
\phi = \frac{2 \pi}{\lambda} \cdot w.
\end{equation}

\noindent Assuming as real the expectation value of the wavefront error given
by an {\XAO} system in the near infrared (i.e. $w \le 10^{-2} \ \mu m$ at $\lambda \sim 1 \ \mu m$),
equation (\ref{eq:Coro-1}) can be approximated as follows:

\begin{equation}\label{eq:Coro-3}
e = p \cdot (1 + i \cdot \phi).
\end{equation}

At this point, the action of an un-specified coronagraph can be formalized directly on the coronagraphic exit pupil plane.
The goal of the coronagraph is to cancel as much as possiblee the amplitude of the electric field along the optical axis
on this plane. Exploiting (\ref{eq:Coro-3}), the resulting on-axis electric field $(e_c)$\ for a perfect
coronagraph\footnote{A perfect coronagraph removes actually the coherent
part of the electric field amplitude due to the on-axis optical beam only, see e.g. \cite{CCe06}; here we consider
the total amplitude in order to simplify the related formalism.} is then:

\begin{equation}\label{eq:Coro-4}
e_c = e - p = i \cdot p \cdot \phi,
\end{equation}

\noindent or, by equation (\ref{eq:Coro-2}), is equal to:

\begin{equation}\label{eq:Coro-5}
e_c = i  \cdot \frac{2 \pi}{\lambda} \cdot p \cdot w.
\end{equation}

\noindent Defining finally $(E_c, P, W)$ as the Fourier transforms (FT) of $(e_c, p, w)$,
equation (\ref{eq:Coro-5}) allows to express the monochromatic post-coronagraphic speckle field $(S)$ as:

\begin{equation}\label{eq:Coro-6}
S(\lambda) \equiv  \left|E_c(\lambda)\right|^2 = \left(\frac{2 \pi}{\lambda}\right)^2 \cdot \left|P \otimes W\right|^2.
\end{equation}

Equation (\ref{eq:Coro-6}) shows that the intensity of a speckle field scales proportionally to $\lambda^{-2}$,
while its chromatic wavelength scaling comes from the fact that the variable involved in the
wavefront $w$ is the spatial frequency $(\nu)$ and not the position $(x)$ in the image plane, i.e.:

\begin{equation}\label{eq:Coro-7}
w(\nu) \equiv FT[W(x)].
\end{equation}

\noindent This indicates that spatial frequency translates into position according to wavelength,
e.g. by applying the standard grating equation as follows:

\begin{equation}\label{eq:Coro-8}
m \cdot \lambda = g \cdot sin(\theta),
\end{equation}

\noindent where $(m)$ is the diffraction order, $(\theta)$ the diffraction angle and $(g)$ is the grating constant
corresponding to the spatial frequency $(\nu)$, or:

\begin{equation}\label{eq:Coro-9}
g(\nu) \equiv \left(\frac{1}{\nu}\right),
\end{equation}

\noindent the position on the image plane returns:

\begin{equation}\label{eq:Coro-10}
x = f \cdot \sin(\theta),
\end{equation}

\noindent $f$ being the focal length of the post-coronagraphic re-imaging optics. Using equations
(\ref{eq:Coro-8}) and (\ref{eq:Coro-9}) this position can be written finally as:

\begin{equation}\label{eq:Coro-11}
x = f \cdot m \cdot \lambda \cdot \nu.
\end{equation}

\begin{figure}[!ht]
\begin{center}
\resizebox{0.48\textwidth}{!}{\includegraphics{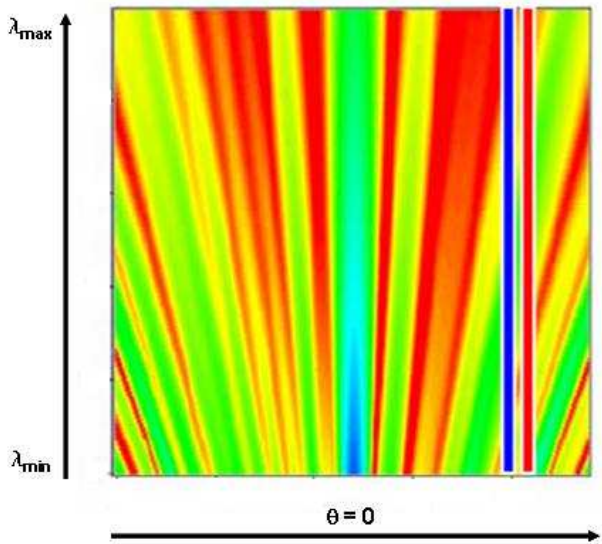}}
\caption{Cuts through the data cube obtained by the chromatical dispersion of a post-coronagraphic speckle pattern resulting from end-to-end simulations of the IFS inside SPHERE (intensity increases from red to blue colors). Position on sky $(\theta)$ is on the horizontal axis, while the spectral range $(\lambda_{min}-\lambda_{max})$ is on the vertical axis. The red and blue lines indicate two spectra taken at different radial distances to the optical axis. Moving along these spectra, a variable pseudo-periodic modulation due to the speckle chromatism is clearly visible (by courtesy of the SPHERE team).}
\label{fg:speckle-spectrum}
\end{center}
\end{figure}

\noindent Equation (\ref{eq:Coro-11}) indicates that the position of a speckle corresponding
to a given fixed spatial frequency due to the post-coronagraphic wavefront error scales linearly
with wavelength \citep{SF02}. More in detail, this means
that for every fixed position in the image plane, speckles corresponding to distinct
spatial frequencies get distinct wavelengths (Figure \ref{fg:speckle-spectrum}).
We call this feature speckle chromatism.

\section{The SDI calibration technique framework}\label{sec:SDI}

In the approach considered in this paper, the fundamental SDI step is the simultaneous
acquisition of images at adjacent wavelengths
in a spectral range where the planetary and stellar spectra differ appreciably.
From ground-based observations, the wavelength bands Y, J, H, and K are well
suited for extrasolar giant planets \citep{BJL08, MC08}, and rocky planets \citep{V08}.

Let $S(\lambda,\theta)$ be the monochromatic spectral signal corresponding to a fixed angular
position $(\theta)$ on sky expressed as the sum of the spectral signal of the star,
$St(\lambda,\theta)$, and the spectral signal of a candidate low-mass companion
(e.g an extrasolar planet) which lies specifically in this angular position, $Pl(\lambda,\theta)$.
Fixing a pair of wavelengths $(\lambda_{1},\lambda_{2})$\ inside the window above, the following relations
hold:

\begin{eqnarray}
S(\lambda_{1},\theta) & = & St(\lambda_{1},\theta)+Pl(\lambda_{1},\theta) \label{eq:SDI-Condition-11} \\
S(\lambda_{2},\theta) & = & St(\lambda_{2},\theta)+Pl(\lambda_{2},\theta) \label{eq:SDI-Condition-12}.
\end{eqnarray}

The basic SDI assumption is that after suitable flux normalization and chromatic re-scaling,
the following relations hold for the boundary wavelengths of the range above:

\begin{eqnarray}
St(\lambda_{1},\theta) & = & St(\lambda_{2},\theta) \label{eq:SDI-Condition-21}\\
Pl(\lambda_{2},\theta) & = & 0 \label{eq:SDI-Condition-22}.
\end{eqnarray}

\noindent Then the difference between $S(\lambda_{1},\theta)$ and $S(\lambda_{2},\theta)$
should return --- in principle --- the spectral signal $Pl(\lambda_{1},\theta)$ only,
i.e. the one appropriate to the low-mass (or extrasolar planet) candidate.
However, while working with narrow-band filters several precautions are required:

\begin{itemize}
\item
an image taken with one filter has to be spatially re-scaled before confronting
it with a second image taken with a different filter due to the speckle scaling described in \S \ \ref{sec:Speckle};
\item
any filter separating two adjacent spectral bands should have similar spectral transmission profiles;
\item
the difference ($\delta\lambda_{ij}$) between the central wavelengths $(\lambda_{i},\lambda_{j})$ of two adjacent filters
should be as small as possible.
\end{itemize}

The last item is the most critical due to the fact that chromatism of the speckle field always induces
a certain amount of phase errors. Adopting the formalism of \cite{M00}, the residual wavefront distortion
can be described through the Fourier transform of the post-coronagraphic wavefront error $(W)$, or by
its relative chromatic phase-error $(\Phi)$.

Adopting the standard approximation for the {\SR} \citep{Mar47}, it is possible to
transfer this RMS wavefront-error $(\sigma_{\Phi})$ on a relative flux variation on the detector plane.
In detail, defining $\sigma_{\Phi}(\lambda)$ as the RMS chromatic wavefront-error, \cite{M00} found the
following relation for the flux residual between images taken with two narrow-band filters $(i,j)$:

\begin{equation}\label{eq:Marois-1}
\frac{\Delta S_{i,j}}{S_{i,j}} = 2 \cdot \sigma_{\Phi}^{2}(\lambda_{i}) \cdot \frac{\delta\lambda_{i,j}}{\lambda_{i}}.
\end{equation}

Equation (\ref{eq:Marois-1}) indicates that with the so called {\em single difference} method the final error
is proportional to:

\begin{itemize}
\item
the variance of the wavefront error: $\sigma_{\Phi}^{2}$,
\item
the relative wavelength separation between the narrow-band filters: $\delta\lambda_{i,j}/\lambda_{i}$.
\end{itemize}

The need for a calibration technique more efficient than SDI but still based on the simultaneous difference
of chromatic images of the same target field was addressed theoretically by \cite{M00}, which showed that the
speckle noise reduction could be much more efficient if observations at three wavelengths were available using
their {\em double difference} method, and tasted experimentally with the discovery the first planet obtained
by using this calibration technique \citep{Lag08}, whose infrared contrast is $> \ 10^{-3.2}$.

Starting from there, it is reasonable to assume that a larger number of images at different wavelengths,
taken with a regular spectral-step, can result in even better reduction of speckle noise with a true
S-SDI calibration technique. The gain could be even larger if observations at several wavelengths would
allow an accurate derivation of the chromatic wavelength scaling, as proposed e.g. by \cite{T07}.
This thought suggests the use of integral field spectroscopy for collecting data simultaneously
at a large number of wavelengths given by the total spectral length and the spectral resolution of a suitable disperser \citep{BAl06}.

Note that such an approach is convenient even in the more conservative approach where modeling of the spectral dependence fail,
simply because a larger number of wavelength pairs can be constructed.

\section{The S-SDI calibration technique framework}\label{sec:S-SDI}

Exploiting an IFU as field stop array over an optical plane conjugated with the focal plane of the telescope itself
allows an appropriate sampling of the post-coronagraphic speckle field defined by equation (\ref{eq:Coro-6}).
The fact that this optical signal gets a finite cut-off spatial frequency proportional to $D/\lambda_{min}$,
where $D$ is the post-coronagraphic pupil size and $\lambda_{min}$ is the spectrograph's cut-on wavelength,
means that a correct spatial sampling on this plane should be imposed searching for suitable sizes
for the separation between adjacent spaxels\footnote{Spaxel indicates a spatial pixel appropriate to the
IFU sub-system inside an IFS-instrument. IFU in turn is the matrix of spaxels which should be placed on
the re-imaged telescope focal plane, working as an optical field-stop array.}, which in turn compose the adopted IFU.
This  sampling condition is detailed in \S \ \ref{sec:-Speckles-Spatial-Sampling}.

The request of a sampling criterion based upon the Shannon theorem is mandatory not
only at the level of the IFU spaxels but also at the level of the detector pixels.
In this case the Shannon sampling condition allows to interpolate correctly, both spatially
and spectrally, the exit slit functions, which in turn are the final output of an integral field spectrograph.
These two sampling conditions are detailed in \S \ \ref{sec:Spatial-Sampling} and
\S \ \ref{sec:Spectral-Sampling} respectively.

\subsection{Spatial sampling of the entrance speckle field}\label{sec:-Speckles-Spatial-Sampling}

Let $F_{in}$ be the focal ratio by which the post-coronagraphic speckle field is projected on the IFU plane.
Theory of image formation (e.g. Goodman 1996) implies then that the cut-off spatial frequency appropriate to $S$
can be written as a function of $F_{in}$ and $\lambda_{min}$ as follows:

\begin{equation}\label{eq:spatial-1}
\nu_C = \left(\frac{1}{F_{in} \cdot \lambda_{min}}\right).
\end{equation}

The spaxel size ($D_{spaxel}$) defines the Nyquist spatial frequency on this plane:

\begin{equation}\label{eq:spatial-2}
\nu_{Ny} \equiv \left(\frac{1}{2 \cdot D_{spaxel}}\right).
\end{equation}

Thus, the Shannon sampling theorem applied to the IFU plane returns:

\begin{equation}\label{eq:spatial-3}
\nu_{Ny} \ge \nu_C.
\end{equation}

\subsection{Spatial sampling of the spectrograph's exit slits}\label{sec:Spatial-Sampling}

The condition avoids aliasing effects when interpolating the array of exit slits
over the whole range of wavelengths considered by the spectrograph, and it may be written through
the following formalism.

The detector pixel size $(d_{pixel})$ defines the Nyquist spatial frequency on this plane:

\begin{equation}\label{eq:super-1}
\mu_{Ny} \equiv \left(\frac{1}{2 \cdot d_{pixel}}\right).
\end{equation}

\noindent Once the final spectrograph's exit slits are imaged on the detector pixels through a fixed output focal ratio $(F_{out})$
and an optical magnification $(m_{IFS})$, theory of image formation, e.g. \cite{{G96}}, implies that their spatial
cut-off frequency is:

\begin{equation}\label{eq:super-2}
\mu_{C} = \left(\frac{1}{\lambda_{min} \cdot m_{IFS} \cdot F_{out}}\right),
\end{equation}

\noindent where $\lambda_{min}$ indicates the shortest wavelength imaged by the spectrograph.
We define the Super-sampling condition as:

\begin{equation}\label{eq:super-3}
\mu_{Ny} \ge \mu_{C}.
\end{equation}

\subsection{Spectral sampling of the speckle field over the entire field of view}\label{sec:Spectral-Sampling}

When working with a speckle pattern data-cube, chromatic re-sampling is needed to obtain
both monochromatic images, as indicated by \cite{M00}, or spectra, as indicated by \cite{T07}.

To this aim \cite{SF02} suggested to adopt a suitable pixel-dependent re-sampling of the speckle field
which varies according to wavelengths, while \cite{RW06} developed a subtraction algorithm based
upon analytical modelings of the spectral content of a speckle field. Anyhow, before any re-sampling recipe,
it is important to find out the exact condition allowing to avoid aliasing errors due to the speckle chromatism effect.

Since the speckle pattern scales proportionally to wavelength (\S \ \ref{sec:Speckle}), a feature located
at an angular distance $\theta$ from the central star at wavelength $\lambda$ moves spectrally at a rate of
$d\lambda/d\theta = \lambda/\theta$. Spatial speckles of width $\delta\theta_S = \lambda/D$ therefore translates
into spectral speckles of width:

\begin{equation}\label{eq:hyper-1}
\delta\lambda_S = \frac{\lambda^2}{\theta \cdot D},
\end{equation}

\noindent i.e., the spectral extension of speckles is inversely proportional to the distance from the field center.
Nyquist sampling of spectral speckles requires spectral sampling ($\delta\lambda_P$) corresponding to half the speckle width,
so far a two-pixel resolving power ($R =\lambda/2\delta\lambda_P$), Nyquist sampling implies the following
condition:

\begin{equation}\label{eq:hyper-2}
R > \frac{\lambda}{\delta\lambda_S} = \theta \cdot \frac{D}{\lambda}.
\end{equation}

\noindent This condition will be fulfilled within a field angle $\theta_{Ny}$,
referred to as the Nyquist radius, given by:

\begin{equation}\label{eq:hyper-3}
\theta_{Ny} = R \cdot \frac{\lambda}{D}.
\end{equation}

We note that it is possible to ensure Nyquist sampling in a system which does not fulfil the Super-sampling
condition written in equation (\ref{eq:super-3}), as long as its field of view does not exceed the Nyquist radius
and as long as the source itself does not contain spectral features which violate the Shannon theorem.
For example, an instrument operating on an $8$ meter telescope at $1.6 \ \mu m$ with a full field of view of $5$
arc-seconds, would require a two-pixel resolving power of at least $60$.
For systems where larger field of view or lower resolving power is required, the Super-sampling condition must be fulfilled.
In these systems, the zone lying within the Nyquist radius fulfils both equation (\ref{eq:super-3})
and and equation (\ref{eq:hyper-2}). We refer to this double fulfilment as Hyper-sampling.

For an integral field spectrograph covering a spectral range fixed between a cut-on ($\lambda_{min}$)
and a cut-off wavelength ($\lambda_{max}$), where $\lambda_c$ represents the central one,
the Hyper-sampling condition will be valid over the whole spectral range within the radius:

\begin{equation}\label{eq:hyper-4}
\theta_{Ny} =\frac{\lambda_{min}^2 \cdot R}{D \cdot \lambda_c}.
\end{equation}

\section{Options for the IFS concept realizing S-SDI}\label{sec:IFS-Options}

IFS needs a very large number of pixels at the level of the final image plane where
the matrix of spectra is acquired by the detector. This issue is particularly important
when spectral and spatial information are recorded simultaneously in the detector plane,
such as for IFS based on the image slicer or the TIGER\footnote{"TIGER" is a French acronym
standing for "Traitement Int\'egral des Galaxies par l'\'Etude de leurs Rays", has \cite{B95}
named their lenslet-based IFS.} concepts.

The image slicer option is more efficient in terms of detector pixels usage, since no separation
between spectra from adjacent pixels is required in one space dimension.
Assuming a square detector, the number of
detector pixels $(N_{det}^2)$ required for a given number of spaxels $(N_{spaxel}^2)$ and number of
spectral samples $(N_{spec})$, is given by the following relation:

\begin{equation}
N_{det}^2 = N_{spaxel}^2 \cdot N_{spec}.
\end{equation}

\begin{figure}[!ht]
\begin{center}
\resizebox{0.40\textwidth}{!}{\includegraphics{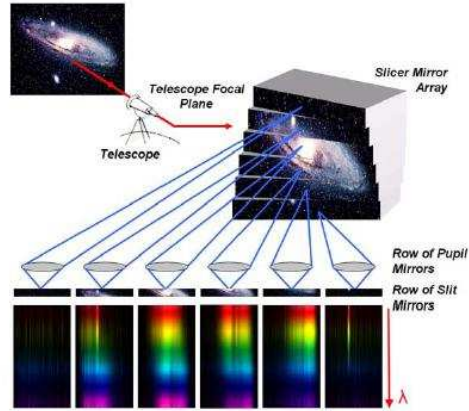}}
\caption{The principle of the image slicer IFU, as figured in Prieto \& Viv\`es 2006
(by courtesy of Dr. Eric Prieto and Dr. S\'ebastien Vives).}
\label{fg:SLICER}
\end{center}
\end{figure}

In this concept a bi-dimensional field of view is divided by mirrors into strips, and
then re-formatted on a mono-dimensional pseudo long-slit (see Figure \ref{fg:SLICER}).
Monochromatic exit slits will be then obtained downstream, by using a standard collimator,
disperser and camera optical system.
A potential problem of the image slicer design concerns the non common path
aberrations in adjacent spaxels of the field of view that fall on different slices.
However this concept has been proved able to obtain (moderately) high-contrast images from
ground even without coronagraphic devices and with moderate {\SR}s $(\sim 0.3-0.5)$
\citep{T07}. A further examination of an image slicer instrument dedicated to high-contrast
diffraction-limited imaging spectroscopy is on progress within the feasibility study
for the future E-ELT Planet Finder facility \citep{KA08}.

On the other hand, non-common path aberrations are expected to be very small
in the case of the TIGER-type concept \citep{B95}, which uses an IFU based on a matrix of lenses
with fixed lens pitch. In this case spectra given by individual spaxels should be separated
on the detector. For a separation of $N_{sep}$ between spectral samples, the required number
of detector pixels becomes:

\begin{equation}
N_{det}^2 = N_{spaxel}^2 \cdot N_{spec} \cdot N_{sep}.
\end{equation}

\begin{figure}[!ht]
\begin{center}
\resizebox{0.45\textwidth}{!}{\includegraphics{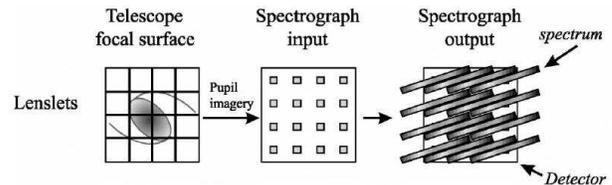}}
\caption{The principle the TIGER IFU as figured by Lee et al. 2001
(by courtesy of Dr. Jeremy Allington-Smith and his co-authors).}
\label{fg:TIGER}
\end{center}
\end{figure}

The lenslet-based concept then requires a large number of detector pixels. However,
the format of image slicer IFS-data on the detector is suited for spectra with many spectral elements,
i.e. $> 10^2$, and relatively small number of spaxels, i.e. $< 10^4$. This are not typical values
for instruments dedicated to planet search that generally requires short spectra ($\sim 20-30$
spectral elements) for a large number of spaxels $(\sim 10^5)$. In order to adequately exploit
the detector, the number of slices should be roughly given by the ratio between the spaxels and
the length of the spectra. This value is $\sim 10^3$ for an integral field spectrograph tuned to planet finding,
which would result in an extremely long pseudo-slit. The format of the image slicer IFU then exacerbates the
problems related to non common paths: indeed photons from adjacent spaxels may have very different
paths through the instrument. It is then difficult to maintain small the phase errors,
possibly compromising most demanding high-contrast imaging.

Given the difficulties inherent to the image slicer solution, we carefully examined the properties
of the lenslet-based design, trying to minimize the separation between spectral samples. To this
aim, we developed the new optical concept proposed by \cite{D06}: BIGRE\footnote{"Bigre" was
the first word uttered by G. Courtes - the inventor of the TIGER concept - while the Authors
explained him all the problems of diffraction-limited IFS and their possible resolution
using this new optical concept. "Bigre" is a French exclamation with a meaning similar to the
British: "Bligh-me" or the Italian: "Accidenti".}. The properties of this design are discussed and
compared to the TIGER ones, starting from \S \ \ref{sec:BIGRE}.

\section{Incoherent and coherent cross-talks of a lenslet-based IFU}\label{sec:CT}

Adopting the formalism of \cite{G96}, any spaxel of an IFU is a sum of linear optical systems.
In the specific case of a lenslet-based IFU these systems are the single lenses.
The coherent and incoherent part of the electric field incoming onto these optical linear systems
are transmitted in a different way through two adjacent spaxels. Specifically, when the
illumination is coherent, the linear responses of adjacent spaxels vary in unison, and therefore
their signals, once transmitted and re-imaged on the spectrograph's slits plane, must be added in complex
amplitude. Contrarily, when the illumination is incoherent, the linear responses of two adjacent
spaxels are statistically independent. This means that their signals, once transmitted and
re-imaged on the spectrograph's slit plane, must be added in intensity.

\begin{figure}[!ht]
\begin{center}
\resizebox{0.40\textwidth}{!}{\includegraphics{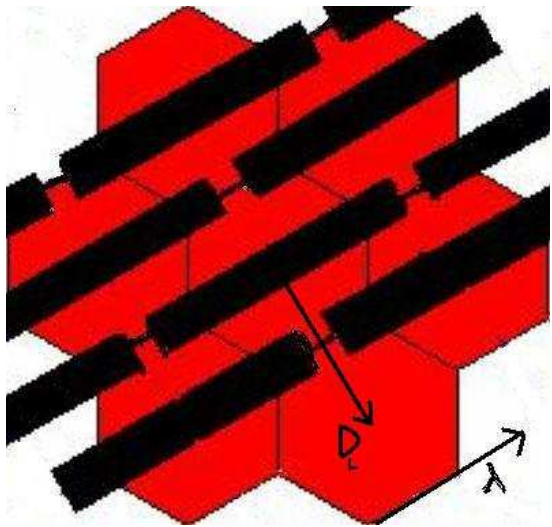}}
\caption{Sketch of the final spectra (black rectangles) superimposed on an array of 7 (red) hexagonal spaxels,
down of a lenslet-based IFU. $D_L$ indicates the IFU lens pitch, while the dispersion axis
is indicated through a black arrow labeled with the symbol $\lambda$.}
\label{fg:zoomed-image}
\end{center}
\end{figure}

Hence, once dispersed and re-imaged by the spectrograph's optics\footnote{The dispersion axis can be defined
orienting the spectrograph's disperser with respect to a reference frame fixed on to the IFU.}, monochromatic slits
corresponding to adjacent spaxels will suffer from a certain amount of interference. We call
this quantity coherent cross-talk. Furthermore, monochromatic slits will be affected by a
spurious amount of signal due to its adjacent spectra. We call this quantity incoherent crosstalk.
With reference to Figure \ref{fg:zoomed-image}, coherent cross-talk is the interference signal
between monochromatic spectrograph's entrance slits which correspond to adjacent lenses, i.e.
separated by a distance equal to the IFU lens pitch\footnote{The pitch of an array of spaxels
is defined as the center-to-center distance among adjacent ones. For a filling factor close to
unity this quantity equals the size of the single spaxel.}. While, incoherent cross-talk is the
spurious signal registered over a fixed monochromatic spectrograph's exit slit and due to its closest spectra,
even if due to photons of different wavelength.

Incoherent and coherent cross-talks represent a major issue identifying the best solution for
the spaxels shape (circular, square, etc...), the lenslet lattice configuration (hexagonal,
square, etc...), and for the geometric allocation of the spectra at the level of the detector
plane. In fact, incoherent and coherent cross-talks are spurious signals --- not removed by the
application of Super- and Hyper-sampling criteria --- which still affect the final array of
spectra, thus damaging the final three-dimensional data cube. The selection of the kind
of field unit to be mounted at the entrance of a lenslet-based integral field spectrograph should then depend on the
estimate of the level of incoherent and coherent signals over the individual exit slits of
such a spectrograph. Additional considerations should enter in this choice, e.g.
the fact that the relevance of the cross-talk terms depends on the wavefront errors after the
coronagraph or that minimization of the cross-talk might result in a system design which is
potentially less efficient when observations are limited by photon noise. In general,
cross-talk should be specified so that its contribution to the contrast error budget is less than the
flat field errors and all remaining spurious effects affecting the the post-coronagraphic
speckle field.

\subsection{Coherent cross-talk: the formalism}\label{sec:CCT}

Basically, coherent cross-talk is the interference of a beam passing through a number of apertures
(individual lenslets) and measured on a screen (the spectrograph's entrance slits plane) conjugated to the detector plane.

Let us assume a flat wavefront impinging onto the IFU lenses. Let now be $E_1$ the complex
electric field of the coherent signal transmitted by spaxel $1$ on the spectrograph's entrance slits plane.
Let be $dE_2$ the stray part of the complex electric field of the coherent signal transmitted by spaxel $2$
(spaxel $2$ being adjacent to spaxel $1$) and evaluated in the position of the slit corresponding to spaxel $1$.
$E_1$ and $dE_2$ are complex quantities that differ according to the phase difference, which is due to
different optical paths through the different apertures (lenslets). The effective coherent intensity
measured on the spectrograph's entrance slits plane and corresponding to the position of spaxel $1$ will then be:

\begin{equation}\label{eq:coherent-1}
I_{1}^{C} \equiv \left|E_1+dE_2\right|^2.
\end{equation}

In the worst case, the phase difference of waves passing through adjacent lenses
is $\pi k$ $(k \in \mathbb{Z})$. In this case and neglecting the term $\left|dE_2\right|^2$ in the
binomial expression of equation (\ref{eq:coherent-1}), the effective coherent intensity proper to
spaxel $1$ becomes:

\begin{equation}\label{eq:coherent-2}
I_{1}^{C} = I_1 + 2 \cdot \left|E_1\right| \cdot \left|dE_2\right| = I_1 \cdot (1+CCT).
\end{equation}

CCT is defined as the coherent cross-talk coefficient:

\begin{equation}\label{eq:coherent-3}
CCT \equiv 2 \cdot \frac{\left|dE_2\right|}{\left|E_1\right|} = 2 \cdot \left(\frac{dI}{I_1}\right)^{1/2},
\end{equation}

\noindent where the stray coherent intensity proper to spaxel $2$ evaluated in the position
of spaxel $1$ is defined as:

\begin{equation}\label{eq:coherent-4}
dI \equiv \left|dE_2\right|^2,
\end{equation}

\noindent and the own coherent intensity of spaxel $1$ is defined as:

\begin{equation}\label{eq:coherent-5}
I_1 \equiv  \left|E_1\right|^2 \label.
\end{equation}

CCT represents the maximum extra-amount of coherent signal on the slit function corresponding
to a fixed lenslet aperture, and its estimate can be given by measuring the square root of the coherent
intensity proper to the slit function corresponding to the adjacent aperture.
However, the total amount of coherent cross-talk is obtained only by adding the
contribution due to all the apertures in the lenslet-array.

\subsection{Incoherent cross-talk: the formalism}\label{sec:ICT}

The amount of spurious incoherent light can be evaluated directly on the detector plane, where a single exit slit
appears as a spectrum. As indicated in Figure \ref{fg:zoomed-image}, any final spectrum is surrounded by several
adjacent spectra.

Let be $I_1(\lambda)$ the intensity proper to a fixed monochromatic exit slit;
due to the presence of an adjacent exit slit its effective incoherent intensity will be:

\begin{equation}\label{eq:incoherent-1}
I^{INC}_1(\lambda) \equiv I_1(\lambda) + dI_2(\lambda),
\end{equation}

\noindent where $dI_2(\lambda)$ is the stray incoherent monochromatic intensity of a given adjacent exit slit,
evaluated at a distance equal to the separation to the fixed one:

\begin{equation}\label{eq:incoherent-2}
dI_2(\lambda) \equiv ICT(\lambda) \cdot I_1(\lambda)
\end{equation}

\noindent where $ICT(\lambda)$ is defined as the monochromatic term of the incoherent cross-talk coefficient (ICT).

The incoherent cross-talk coefficient corresponding to the spectrograph's
wavelengths range $(\lambda_{min} - \lambda_{max})$ is then defined as:

\begin{equation}\label{eq:incoherent-3}
ICT \equiv \int_{\lambda_{min}}^{\lambda_{max}}{\left(\frac{I^{INC}_1(\lambda) - I_1(\lambda)}{I_1(\lambda)}\right) \ d \lambda}.
\end{equation}
 	
Thus --- differently to the coherent case --- the incoherent cross-talk must be considered on the
detector plane, searching for spectral alignments for which the distance among adjacent spectra is minimized.
Once this spectral alignment is found, an estimate of ICT can be given by measuring the incoherent intensity
of a single monochromatic exit slit at the distance equal to the transversal separation among adjacent spectra.
However, the total amount of incoherent cross-talk is obtained only by adding the contribution of all the spectra
imaged onto the detector plane.

\section{Diffraction-limited integral field spectroscopy with the TIGER concept}\label{sec:TIGER}

In classical TIGER design optimized for seeing limited conditions the spaxels (or microlenses) composing
the IFU are much bigger than the Airy disk, providing therefore resolved images of the telescope entrance pupil,
which in turn represent the entrance slits of this kind of integral field instrument, see e.g. \cite{B95, B01}.

Differently, in the case of high-contrast imaging the microlenses sample the telescope image
according to the Shannon theorem. Each microlens acts like a diaphragm isolating a portion of the
incoming electric field and concentrates it into a micropupil image in the focal plane of the microlens,
acting as the entrance slit function of the spectrograph. The micropupil image is the convolution between
the geometrical pupil image and the PSF of the microlens. As seen below, Nyquist sampling
of the focal plane implies that the telescope entrance pupil is unresolved by the microlens.

For a circular lens of diameter $D_{spaxel}$ the transmission function is $\Pi(u/D_{spaxel})$,
where $u$ is the image co-ordinate normalized to the lens diameter and $\Pi(x)$ is a top-hat function
with unitary transmission within the unitary diameter and zero outside of this diameter.
According to condition (\ref{eq:spatial-3}), the size of the single microlens should be:

\begin{equation}\label{eq:tiger-ifu-1}
D_{spaxel} \le  \left(\frac{F_{in} \cdot \lambda_{min}}{2}\right).
\end{equation}

\noindent Following \cite{BW65}, the monochromatic full-width-half-maximum (FWHM) of the PSF proper
to a circular microlens with focal length $f_{out}$ is:

\begin{equation}\label{eq:tiger-ifu-2}
FWHM = f_{out} \cdot 1.02 \cdot \frac{\lambda}{D_{spaxel}},
\end{equation}

\noindent while the geometrical diameter of the micropupil is:

\begin{equation}\label{eq:tiger-ifu-3}
D_{MPG} = \frac{f_{out}}{F_{in}}.
\end{equation}

\noindent Combining equations (\ref{eq:tiger-ifu-1}), (\ref{eq:tiger-ifu-2}) and (\ref{eq:tiger-ifu-3})
we obtain then:

\begin{equation}\label{eq:tiger-ifu-4}
FWHM =  2.04 \cdot \left(\frac{D_{MPG} \cdot \lambda}{\lambda_{min}}\right).
\end{equation}

This size is therefore at least twice as wide as the geometrical pupil, and so the convolution
product is approximately equal to the microlens PSF.

Thus, we can stay that the field distribution onto the spectrograph's slit plane approximates the one
proper to an unresolved micropupil, which is described by the Jinc function\footnote{We define
Jinc function as the Fourier transform of circular aperture: $Jinc(x) = \left(2 \cdot J_1(\pi \cdot x)\right)/\left(\pi \cdot x\right)$, where $J_1$ indicates the Bessel-J function of order one.} corresponding to the microlens aperture:

\begin{equation}\label{eq:tiger-ifu-5}
e_{pupil}(s) \sim Jinc(s),
\end{equation}

\noindent $s$ being defined as the pupil the co-ordinate normalized to $\lambda \cdot F_{out}$,
where $F_{out}$ is:

\begin{equation}\label{eq:tiger-ifu-5-1}
F_{out} \equiv \frac{f_{out}}{D_{spaxel}}
\end{equation}

Finally, the slit function will be the square modulus of this signal:

\begin{equation}\label{eq:tiger-ifu-6}
SF(s) = \left|e_{pupil}(s)\right|^2.
\end{equation}

\begin{figure}[!ht]
\begin{center}
\resizebox{0.45\textwidth}{!}{\includegraphics{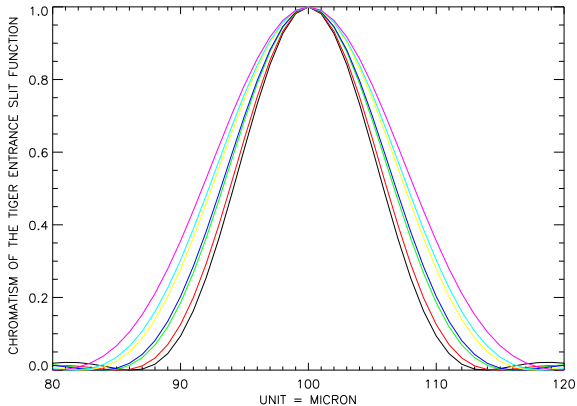}}
\caption{Normalized TIGER entrance slit function: the example shows the case of the IFU optimized for SPHERE
in the working wavelengths range: $0.95-1.35 \ \mu m$;
colors indicate seven distinct wavelengths.}
\label{fg:TIGER-LSF}
\end{center}
\end{figure}

\subsection{Sampling analysis applied to the TIGER concept}\label{sec:TIGER-IFS-equations}

As indicated by equation (\ref{eq:tiger-ifu-6}), the single spectrograph's slit is an un-bound signal
whose size varies linearly with wavelength. The final pixel size defines the spatial Nyquist
frequency on the spectrograph image plane according equation (\ref{eq:super-1}). Due to its un-bound nature,
the spatial cut-off frequency of the spectrograph's exit slit gets the finite value fixed by equation (\ref{eq:super-2}).
Then, following condition (\ref{eq:super-3}), Super-sampling imposes a lower limit to the output
focal ratio by which the single microlens generates its corresponding micropupil:

\begin{equation}\label{eq:tiger-ifu-7}
F_{out} \ge \left(\frac{2 \cdot d_{pixel}}{\lambda_{min} \cdot m_{IFS}}\right).
\end{equation}

Output focal ratios lower than the one fixed by equation (\ref{eq:tiger-ifu-7})
introduce aliasing errors in the sampled spectrum, unless the field is smaller than the Nyquist radius.
According to condition (\ref{eq:hyper-4}), this latter depends on the post-coronagraphic pupil size,
the spectrograph's working wavelengths range and its spectral resolution.
Hence, the true Hyper-sampling is obtained when this radius matches with the maximum image field radius,
which in turn is related to the spectrograph's resolving power. Then, for a fixed resolving power,
Hyper-sampling is then a matter of allocation of the array of final spectra onto the detector pixels,
which in turn depends on the accepted cross-talk levels.

\begin{table}[h]
\caption{Independent parameters of a TIGER-oriented IFS}
\begin{center}
\begin{tabular}{ll}
Post-coronagraphic pupil size                       &  $ \equiv D            $\\
IFS cut-on wavelength                               &  $ \equiv \lambda_{min}$\\
Size of the single TIGER microlens                  &  $ \equiv D_{spaxel}   $\\
Focal length of the TIGER microlens                 &  $ \equiv f_{out}      $\\
IFS detector pixel size                             &  $ \equiv d_{pixel}    $\\
IFS optical magnification                           &  $ \equiv m_{IFS}      $\\
IFS disperser (2-pixel) resolving power             &  $ \equiv R            $\\
\end{tabular}
\end{center}
\label{tab:tiger-ifs-parameters}
\end{table}

\section{Diffraction-limited integral field spectroscopy with the BIGRE concept}\label{sec:BIGRE}

Cross-talk in diffraction-limited TIGER-oriented IFU is generally quite large
because the output slit functions, taking the form of an Airy pattern, decrease
slowly with distance from the center. Suitable apodization of the microlenses might
in principle be used to reduce the cross-talk terms, but the feasibility of such
a scheme remains to be demonstrated. We consider instead an alternative lenslet-based
optical scheme that we call BIGRE, which we consider to be much more practical.

As in the TIGER case, the BIGRE spaxel consist of a microlens which acts essentially
as a diaphragm isolating a portion of the incident electric field. This lens, of focal
length $f_1$, focalizes the field into an unresolved micropupil with a field distribution
described by equation (\ref{eq:tiger-ifu-5}).
Differently to the TIGER case, we place a second microlens at a distance equal
to its focal length $f_2$, behind the micropupil. This lens collects field and reproduces
an image of the first lens, behind the micropupil. When $f_2 < f_1$, the final image
is reduced, resulting in the same flux-concentrating effect as in the original TIGER concept,
but without the field-pupil inversion. We define $K$ factor as the spaxel de-magnification factor:

\begin{equation}\label{eq:bigre-ifu-1}
K \equiv \frac{f_1}{f_2}.
\end{equation}

Ideally, for infinitely wide optics throughout the following spectrograph, the slit function is
a perfectly bound top-hat function, so no cross-talk would be present between spaxels.
Of course, this is not physical, and the following finite sized optics modifies the slit function as we
will see in the following.

It may also be argued that a perfect top-hat function is not the ideal slit function from a sampling
point of view, since its modulus transfer function (MTF) will be un-bound and create some aliasing.
As we will see, the implementation of a diaphragm of appropriate size, modifies the slit function in
a way which turns out to be beneficial both from a cross-talk and from a sampling point of view.

\begin{figure}[!ht]
\begin{center}
\resizebox{0.45\textwidth}{!}{\includegraphics{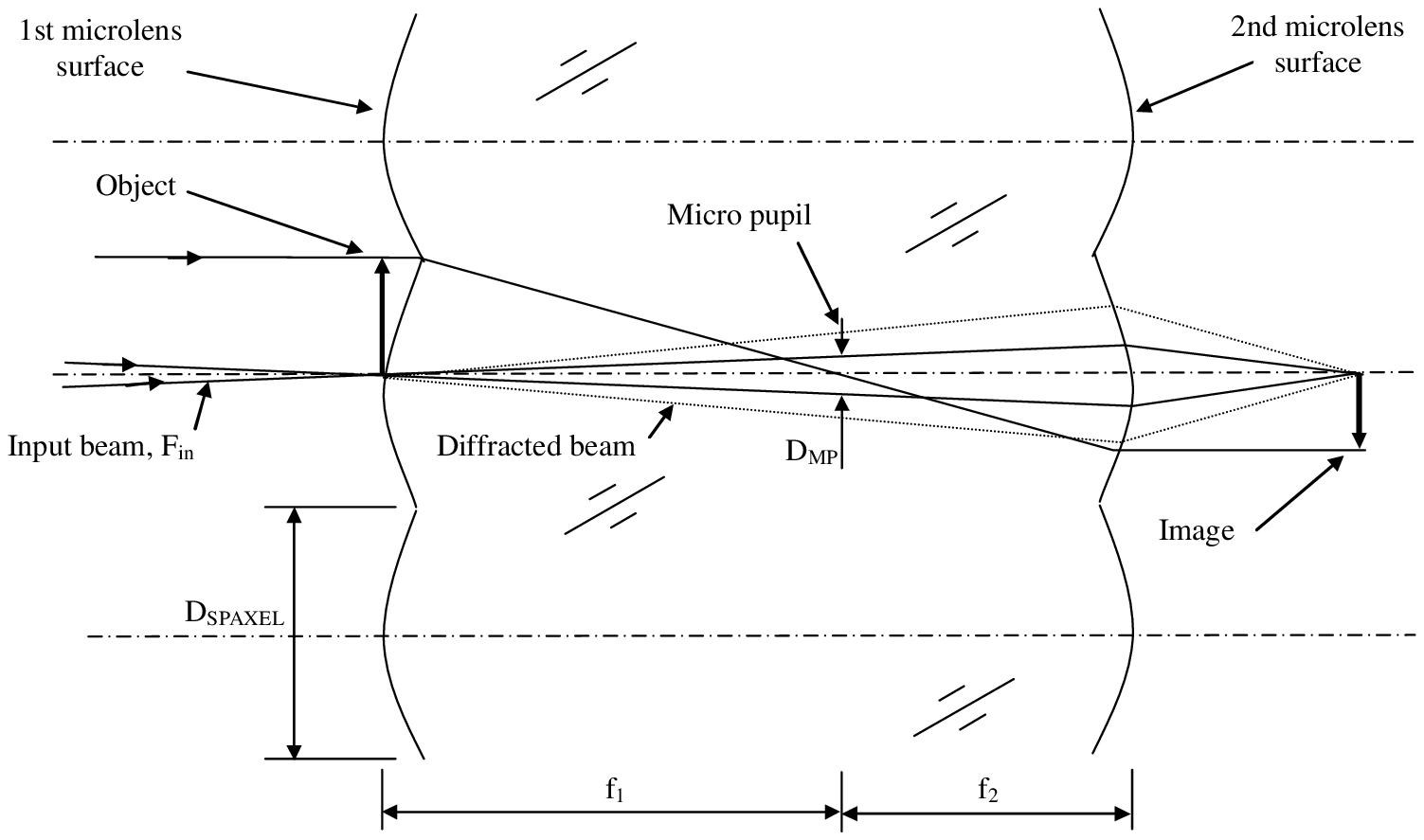}}
\caption{Scheme of a BIGRE spaxel working at the diffraction limit with an un-resolved entrance pupil. The first lens lies on a focal plane and re-images a micropupil at its focal distance ($f_1$). The electric field imaged onto this optical plane is a sinc function (one dimension) or a Jinc function (two dimensions). This signal is filtered by a top-hat transmission function and finally re-imaged onto an image plane by the second lens. The distance between this intermediate pupil plane and the second lens is its focal length ($f_2$). The electric field imaged by this second lens is an un-bound signal with upper envelope much steeper than the one proper to a sinc profile ($\propto u^{-1}$) or a Jinc profile ($\propto u^{-1.5}$).}
\label{fg:BIGRE-IFU}
\end{center}
\end{figure}

Figure \ref{fg:BIGRE-IFU} shows the BIGRE spaxel conceptually, indicating its dimensions and the geometrical
ray paths. The two lenslet-array are implemented as the two surfaces of a single component and the micro-pupil
array occurs within the component. In principle, it would be possible to implement a mask in this micro-pupil
image, but this option has not been retained in view of complexity of manufacturing and aligning a system of three
micro-optical elements (lens, diaphragm, lens). Instead, we consider the second lens and the subsequent collimation
optics to be sufficiently large to not significantly modify the field transmission, implementing the mask in the
metapupil image formed onto the spectrograph's dispersion element,
see Figure \ref{fg:IFS-optical-concept}.

While the geometrical micropupil size is given by the focal ratio of the input
beam according to equation (\ref{eq:tiger-ifu-3}), the characteristic size of the diffractive
micropupil is:

\begin{equation}\label{eq:bigre-ifu-2}
D_{MP} = \lambda \cdot \frac{f_1}{D_{spaxel}}.
\end{equation}

In the following, we use a pupil co-ordinate unit, $s$, which is normalized to $D_{MP}$,
allowing us to discuss the size of the pupil diaphragm without worrying about the optical design
characteristics of the intervening optics.
For the above assumption concerning relatively undisturbed propagation of the electric field from
the micropupils to the spectrograph pupil, we need to ensure that the diffracted beam does not get
truncated by the second microlens edge. For this, a criterion would be to make sure the diffractive
micropupil is much smaller than the spaxel diameter: $D_{MP} << D_{spaxel}$. Plugging this condition
into equation (\ref{eq:bigre-ifu-2}), we get the following condition on the focal length of the first surface:

\begin{equation}\label{eq:bigre-ifu-3}
f_1 << \frac{D_{spaxel}^2}{\lambda}.
\end{equation}

Introducing a pupil mask defined by:

\begin{equation}\label{eq:bigre-ifu-4}
PM(s) \equiv \Pi(s/S_{PM}),
\end{equation}

\noindent where $\Pi(x)$ is a top-hat function with unitary transmission within the diameter $S_{PM}$,
which is turn is the pupil mask size in units of $s$. We can express the electric field distribution
in the exit slit plane as:

\begin{equation}\label{eq:bigre-ifu-5}
E_{slit}(u) = FT\left[e_{pupil(s)} \cdot PM(s)\right].
\end{equation}

Hence, evoking the convolution theorem and remembering that the field in the pupil plane is the Fourier
transform of the field in the spaxel, this can be re-written as:

\begin{equation}\label{eq:bigre-ifu-6}
E_{slit}(u) = \Pi(u/D_{spaxel}) \otimes FT\left[\Pi(s/S_{PM})\right],
\end{equation}

\noindent i.e. the convolution between a top-hat function corresponding to the original spaxel
transmission function and a Jinc function corresponding to the micropupil mask.

Finally, the slit function is the square modulus of this signal:

\begin{equation}\label{eq:bigre-ifu-7}
SF(u) = \left|E_{slit}(u)\right|^2,
\end{equation}

\noindent and the its spectral modulation transfer function is:

\begin{equation}\label{eq:bigre-ifu-8}
MFT(s) \equiv \left|FT\left[SF(u)\right]\right|.
\end{equation}

\begin{figure}[!ht]
\begin{center}
\resizebox{0.45\textwidth}{!}{\includegraphics{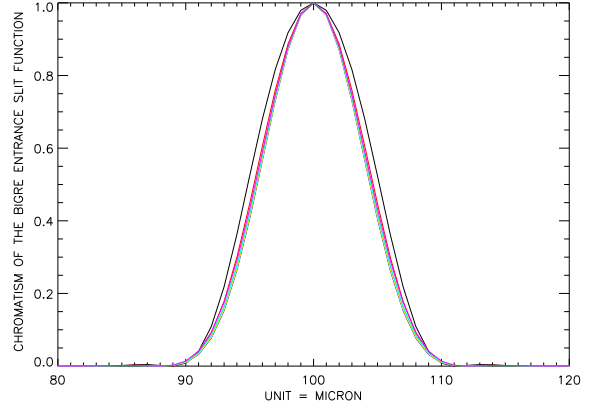}}
\caption{Normalized BIGRE entrance slit function: the example shows the case of the IFU optimized for SPHERE
in the working wavelengths range: $0.95-1.35 \ \mu m$;
colors indicate seven distinct wavelengths.}
\label{fg:BIGRE-LSF}
\end{center}
\end{figure}

\subsection{Sampling analysis applied to the BIGRE concept}\label{sec:BIGRE-IFS-equations}

According to condition (\ref{eq:spatial-3}), the input focal ratio of the light coming to the
single BIGRE spaxel should be:

\begin{equation}\label{eq:bigre-ifu-9}
F_{in} \ge \left(\frac{2 \cdot D_{spaxel}}{\lambda_{min}}\right).
\end{equation}

From the paraxial perspective, input and output focal ratios of a BIGRE spaxel are
related through the $K$ factor as follows:

\begin{equation}\label{eq:bigre-ifu-10}
F_{out}^{G} = \frac{F_{in}}{K},
\end{equation}

\noindent and the geometric micropupil size returns:

\begin{equation}\label{eq:bigre-ifu-11}
D_{MPG} \equiv \frac{f_1}{F_{in}}.
\end{equation}

From the diffractive perspective, the output focal ratio is fixed only when the size of
the pupil mask is fixed on the micropupil plane, due to the un-bound nature of this micropupil profile.
The characteristic size of the diffractive micropupil ($D_{MP}$) can be parameterized in terms of the
focal ratio of the first BIGRE lens ($F_1$) and the spectrograph central wavelength ($\lambda_c$)
as follows:

\begin{equation}\label{eq:bigre-ifu-12}
D_{MP} = S_{PM} \cdot \lambda_c \cdot F_1.
\end{equation}

The diffractive output focal ratio ($F_{out}$) results then from the following equation:

\begin{equation}\label{eq:bigre-ifu-13}
F_{out} = F_{out}^G \cdot \frac{D_{MPG}}{D_{MP}},
\end{equation}

\noindent or, exploiting equations (\ref{eq:bigre-ifu-10}), (\ref{eq:bigre-ifu-11}) and (\ref{eq:bigre-ifu-12}):

\begin{equation}\label{eq:bigre-ifu-14}
F_{out} = \left(\frac{D_{spaxel}}{K \cdot S_{PM} \cdot \lambda_c}\right).
\end{equation}

Finally, by equations (\ref{eq:super-2}) and (\ref{eq:bigre-ifu-14}),
the spatial cut-off frequency of the spectrograph's exit slit becomes:

\begin{equation}\label{eq:bigre-ifu-15}
\mu_{C} = \left(\frac{K \cdot S_{PM}}{D_{spaxel} \cdot m_{IFS}}\right).
\end{equation}

Equation (\ref{eq:bigre-ifu-15}) indicates that the actual profile of the spectrograph's slit function
is no longer a bound signal, just because the pupil mask gets a finite size. The actual size
of the final exit slit function will be then an un-bound signal with spatial cut-off frequency depending
both on the size of this pupil mask, on the de-magnification factor of the BIGRE spaxel
and the magnification of the re-imaging optics. By this analysis, Super-sampling applies to the final exit slit
function through condition (\ref{eq:super-3}) as follows:

\begin{equation}\label{eq:bigre-ifu-16}
2 \cdot d_{pixel} \le \left(\frac{D_{spaxel} \cdot m_{IFS}}{K \cdot S_{PM}}\right).
\end{equation}

We can now study the effect of varying the pupil mask size on the slit function in term of cross-talk
performance and on the MTF in term of aliasing.

Choosing a very large pupil mask, $S_{PM} >> 1$, corresponds to transmitting the spaxel transmission
profile without modification: its FWHM is $D_{spaxel}$ and cross-talk is zero. The MTF is a Jinc function
with first zero at $1.22/s$, so sampling this slit with two pixels across its width causes aliasing of up to
around $15 \%$. On the other hand, choosing a very small pupil mask, $S_{PM} << 1$
creates a wide slit function with a shape approximately equal to an Airy pattern
of FWHM $\sim D_{spaxel}/S_{PM}$. The cross-talk is the same as that found for the TIGER case,
and the MTF is equal to the classical MTF function for diffraction limited optical systems.
Sampling corresponding to half of the FWHM is exempt of any aliasing.

\begin{figure}[!ht]
\begin{center}
\resizebox{0.45\textwidth}{!}{\includegraphics{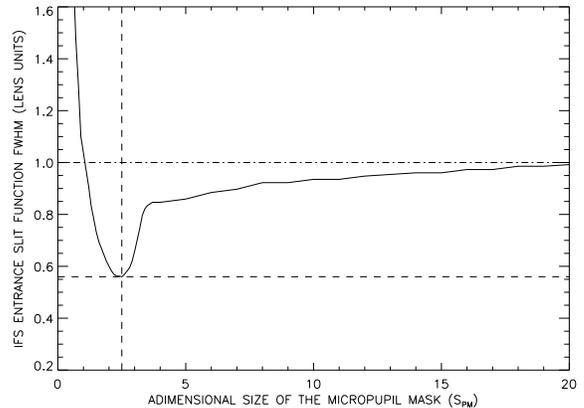}}
\caption{FWHM of the BIGRE entrance slit function profile as a function of the pupil mask size $S_{PM}$.
This FWHM gets its absolute minimum when $S_{PM} = 2.5$. Dot-dashed horizontal line indicates the
asymptotic trend of the FWHM, corresponding to the pupil mask sizes towards the limit: $S_{PM} = \infty$.
Dashed horizontal line indicates FWHM value corresponding to the absolute minimum $S_{PM} = 2.5$,
this one traced with a dashed vertical line.}
\label{fg:BIGRE-FWHM}
\end{center}
\end{figure}

\begin{figure}[!ht]
\begin{center}
\resizebox{0.45\textwidth}{!}{\includegraphics{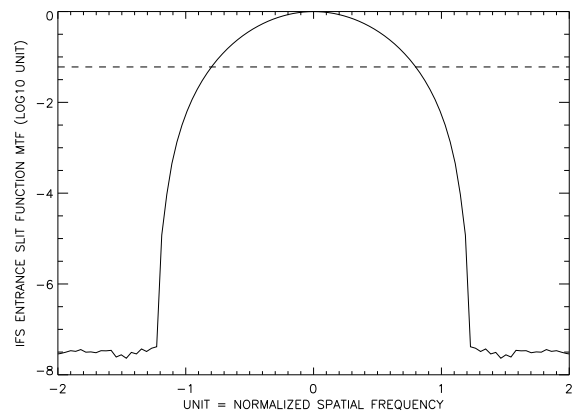}}
\caption{MTF of the BIGRE entrance slit:
filtering out the spatial frequencies above the one corresponding to the entrance slit FWHM
(dashed horizontal line) produces a limited aliasing error: $MTF(s) \ \mid \ s > s_{FWHM} < 6\%$.}
\label{fg:BIGRE-MTF}
\end{center}
\end{figure}

It is somewhat surprising to find in between these two extremes, the evolution of the FWHM is not monotonic, but passes
through a minimum, located at $S_{PM} = 2.5$. At this position, the slit function has
a Gaussian-like bell shape, and its FWHM is $\sim 0.56 \cdot D_{spaxel}$, see Figure \ref{fg:BIGRE-FWHM}.
The slit function falls off rapidly, and its first secondary maximum peaks at values $< 0.001$.
Compared with the ones of an Airy function ($> 0.01$) this ensures a low level of cross-talk.
The spectral MTF also resembles a Gaussian function, with a monotonic fall-off, see Figure \ref{fg:BIGRE-MTF}.
For a sampling of two pixels across the FWHM the aliasing is well below $6\%$, see Figure \ref{fg:BIGRE-MTF}.

The presence of a minimum indicates that the size of the slit function could be stable with respect to variations
in wavelength, indicating that the pupil mask works chromatically as a pupil apodization \citep{JRD64}.
This is indeed the case, as indicated in Figure \ref{fg:BIGRE-LSF}, where the slit function is plotted for
several wavelengths in the range $0.95-1.35 \ \mu m$. We study the wavelength evolution of coherent and incoherent
cross-talks in \S \ \ref{sec:SPHERE-IFS}.

This analysis suggests then that the spectrograph's entrance slit shape can be fixed selecting
properly the pupil mask's minimum size. Once projected on the final detector plane,
Super-Sampling can be fixed by imposing that two pixels cover the spectrograph's exit slit FWHM:

\begin{equation}\label{bigre-ifu-17}
2 \cdot d_{pixel} \le \left(\frac{m_{IFS}}{s_{FWHM}}\right),
\end{equation}

\noindent where $s_{FWHM}$ is the spatial frequency corresponding
to a spatial period equal to the slit function FWHM. While, according condition (\ref{eq:hyper-4}),
Hyper-sampling depends on the post-coronagraphic pupil size, the spectrograph's working
wavelengths range and its resolving power.

\begin{table}[h]
\caption{Independent parameters of a BIGRE-oriented IFS}
\begin{center}
\begin{tabular}{ll}
Post-coronagraphic pupil size                               &  $ \equiv D            $\\
IFS cut-on wavelength                                       &  $ \equiv \lambda_{min}$\\
IFS central wavelength                                      &  $ \equiv \lambda_c    $\\
Size of the single BIGRE microlens                          &  $ \equiv D_{spaxel}   $\\
Focal length of the first BIGRE optical surface             &  $ \equiv f_1          $\\
Focal length of the second BIGRE optical surface            &  $ \equiv f_2          $\\
Size of the pupil mask in unit of $\lambda_c\cdot F_1$      &  $ \equiv S_{MPM}      $\\
IFS detector pixel size                                     &  $ \equiv d_{pixel}    $\\
IFS optical magnification                                   &  $ \equiv m_{IFS}      $\\
IFS disperser (2-pixel) resolving power                     &  $ \equiv R            $\\
\end{tabular}
\end{center}
\label{tab:bigre-ifs-parameters}
\end{table}

\begin{figure}[!ht]
\begin{center}
\resizebox{0.45\textwidth}{!}{\includegraphics{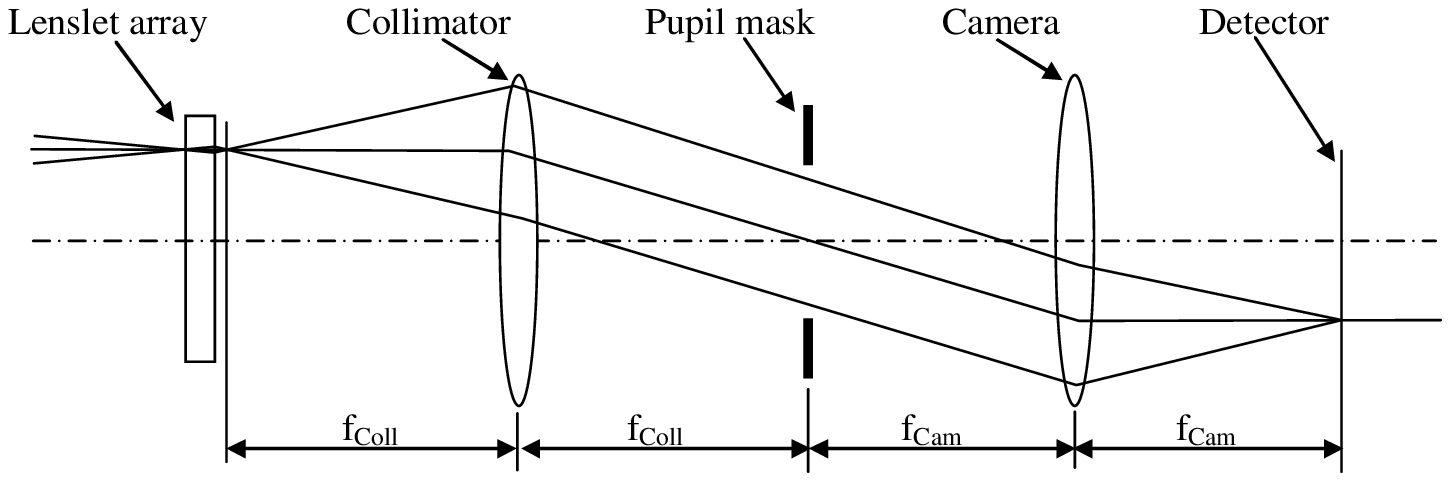}}
\caption{BIGRE spectrograph concept: the entrance slits plane is filled with
the micro-images of the the first surfaces of the BIGRE spaxels  and the spatial
filtering is done in the metapupil plane forming between collimator, having
focal length $f_{Coll}$ and the re-imaging optics, having focal length $f_{Cam} \equiv m_{IFS} \cdot f_{Coll}$.}
\label{fg:IFS-optical-concept}
\end{center}
\end{figure}

Finally, as shown in Figure \ref{fg:IFS-optical-concept}, the aim of the optics downstream the BIGRE lenslet-array
is to re-image the entrance slit into the spectrograph's image plane with the highest stability and optical quality;
for this reason the optical design can be fully dioptric.
The requested stability is assured imposing the telecentricity of the entrance pupil.
In turn, this implies that the metapupil forming between collimator and re-imaging optics,
which is the result of the over-position of individual micropupils forming inside the
lenslet-array, has a size equal to the size of a single micropupil, once
properly magnified by the ratio between the equivalent focal length of the collimator
optics and the focal length of the second optical surface of the single
BIGRE lens. While, the spatial filtering of the micropupils is obtained by adopting
a unique pupil stop placed onto this spectrograph's metapupil plane with a physical size $(D_{PS})$ obtained as follows:

\begin{equation}\label{eq:bigre-ifs-18}
D_{PS} = D_{MP} \cdot \left(\frac{f_{Coll}}{f_2}\right),
\end{equation}

\noindent where $D_{MP}$ is fixed by equation (\ref{eq:bigre-ifu-12}).
Thus, a suitable dispersing device can be inserted in the optical train after this pupil stop
allowing to image the exit slits as true spectra on the spectrograph's image plane.

\section{BIGRE and TIGER IFU solutions for SPHERE IFS}\label{sec:SPHERE-IFS}

Coherent and incoherent cross-talks establish the actual imaging contrast measured onto
the detector $(C_{pixel})$ with respect to the reference value $(C_{spaxel})$ depending
on the spatial sampling of the post-coronagraphic speckle field. Their difference depends
on cross-talk just because the optical signal imaged by a fixed spaxel is spread over
a number of detector pixels larger than the ones corresponding to it by geometrical
optical propagation only, in a way which is proportional to the levels of cross-talks.
When the cross-talk coefficients are sufficiently small this difference can be approximated as:

\begin{equation}\label{eq:confront-1}
C_{pixel} - C_{spaxel} \approx - n \cdot (CCT + ICT),
\end{equation}

\noindent where $n$ is the number of adjacent spaxels around the fixed one,
while CCT and ICT are the cross-talk coefficients defined by equations
(\ref{eq:coherent-3}) and (\ref{eq:incoherent-3}), respectively.

\begin{figure}[!ht]
\begin{center}
\resizebox{0.45\textwidth}{!}{\includegraphics{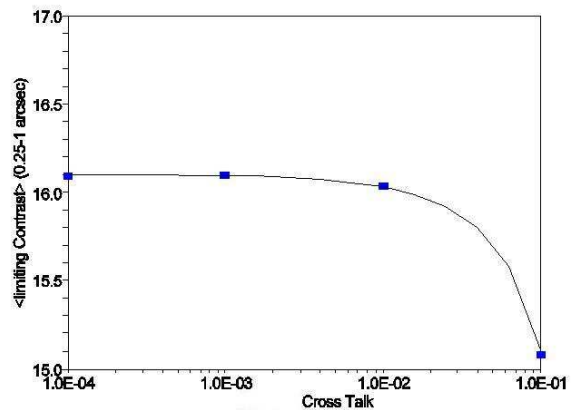}}
\caption{Average 5-$\sigma$ contrast over an azimuthal area comprised between
0.25 and 1 arc-second from a J = 3.75 (mag) star. The simulations are for $10^4 \ sec$ exposure time
and $90 \ degree$ field rotation. Filled squares are results of the IFS end-to-end simulations while the solid line represents the contrast curve expected by exploiting equation (\ref{eq:confront-1}) in order to obtain $C_{pixel}$ via the S-SDI calibration technique
The adopted post-coronagraph contrast profile is the one presented by \cite{Boc08}.}
\label{fg:SPHERE-IFS-CT-CL}
\end{center}
\end{figure}

For the IFS channel of SPHERE the requested cross-talk coefficients have
been determined through a series of simulations devoted to measure the contrast
capabilities of this integral field spectrograph. The result is that the impact of
cross-talk is well reduced when the Super-sampling condition is verified.
This fact can be explained heuristically remembering the meaning of the cross-talk errors
over a fixed spectrograph's exit slit: to replace its monochromatic intensity
with the sum of this intensity and the average of the intensities proper to the exit slits
corresponding to its adjacent spaxels (via the coherent cross-talk coefficient)
together with the average of the intensities proper to the exit slits
corresponding to its adjacent spectra (via the incoherent cross-talk coefficient).
In the case of Super-sampling, adjacent exit slits do not suffer
from a mutual shape variations, instead they suffer only from mutual differences in intensity
due to the input post-coronagraphic speckle field. In this way, the residual between
a fixed exit slit's intensity an its ideal value (free from cross-talk errors) becomes small beyond
a fixed threshold depending on the speckle rejection capabilities of the coronagraph.
No gain in contrast is then possible for further decrements of the cross-talk coefficients.
In the case of the IFS simulations this threshold returns to be 0.01, see Figure \ref{fg:SPHERE-IFS-CT-CL}.
As a conclusion, the IFU solutions for the IFS of SPHERE should be compliant with this specification.

\begin{figure}[!ht]
\begin{center}
\resizebox{0.45\textwidth}{!}{\includegraphics{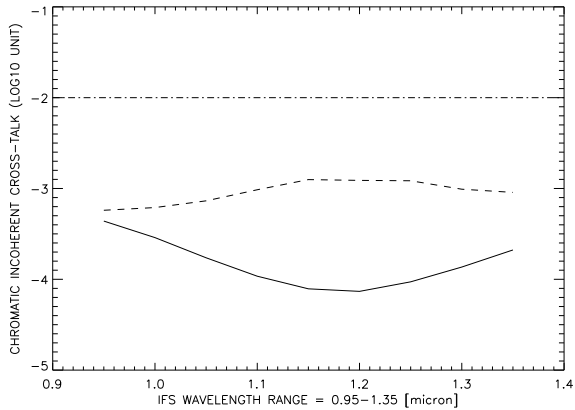}}
\caption{Incoherent cross-talk coefficient as a function of the wavelength in the range
$0.95-1.35 \ \mu m$. Solid line represents the BIGRE solution, dashed line the TIGER one
and dot-dashed line the SPHERE IFS specification.}
\label{fg:BIGREvsTIGER-ICT}
\end{center}
\end{figure}

\begin{figure}[!ht]
\begin{center}
\resizebox{0.45\textwidth}{!}{\includegraphics{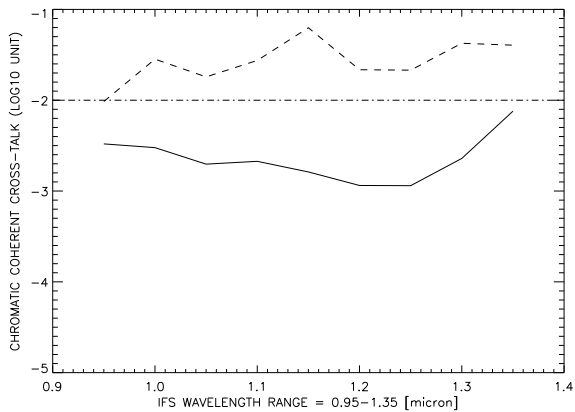}}
\caption{Coherent cross-talk coefficient as a function of the wavelength in the range
$0.95-1.35 \ \mu m$. Solid line represents the BIGRE solution, dashed line the TIGER one
and dot-dashed line the SPHERE IFS specification.}
\label{fg:BIGREvsTIGER-CCT}
\end{center}
\end{figure}

Figures \ref{fg:BIGREvsTIGER-ICT} and \ref{fg:BIGREvsTIGER-CCT} show the levels of incoherent cross-talks,
respectively in the TIGER and BIGRE designs optimized for SPHERE, plotted against wavelength. While the incoherent
cross-talk is below the $1 \%$ threshold for both designs, the BIGRE design is clearly superior, showing a minimum towards
the middle of the range corresponding to the wavelength at which the pupil mask is optimal. The coherent cross-talk is greater
than the incoherent one, as expected, but again the BIGRE design shows superior performance, and remains well below
the $1 \%$ threshold across the spectral range of interest. The TIGER design, on the other hand, is not within the
specified limit.

Table \ref{tab:bigre-ifs-sphere-parameters} resumes the solution we found for the BIGRE-oriented
IFU of SPHERE allowing to reach the requested coherent and incoherent cross-talk levels.
With this solution Hyper-sampling is verified within the whole scientific field of view:
the Nyquist radius is larger than the radial field of view imaged by the spectrograph's optics ($1.25$ arc-seconds).
Table \ref{tab:tiger-ifs-sphere-parameters} summarizes the solution we found for the TIGER-oriented IFU of SPHERE.
This one allows to reach the requested incoherent cross-talk limit but not the requested coherent cross-talk limit,
while Hyper-sampling is well verified as in the previous case.

Based on these results, a BIGRE design is chosen for the IFS channel of SPHERE, configured with circular spaxels
in a hexagonal lattice configuration.

\begin{table}[!ht]
\caption{Basic parameters proper to the BIGRE solution of SPHERE}
\begin{center}
\begin{tabular}{cccc}
\hline
$D_{spaxel} =  161.5 \ \mu m$ & $\lambda_{min} = 0.95 \ \mu m$ & $K = 4.1$ & $F_{out}    =  12$ \\
$m_{IFS}    = 1.69$           & $d_{pixel}     = 18 \ \mu m$   & $R = 54$\\
\hline
\end{tabular}
\end{center}
\label{tab:bigre-ifs-sphere-parameters}
\end{table}

\begin{table}[!ht]
\caption{Basic parameters proper to the TIGER solution of SPHERE}
\begin{center}
\begin{tabular}{ccc}
\hline
$D_{spaxel} =  150 \ \mu m$ & $\lambda_{min} = 0.95 \ \mu m$     & $F_{out} = 7$\\
$m_{IFS}    =  2.4$         & $d_{pixel}     = 18    \ \mu m $   & $R       = 24$\\
\hline
\end{tabular}
\end{center}
\label{tab:tiger-ifs-sphere-parameters}
\end{table}

\begin{figure}[!ht]
\begin{center}
\resizebox{0.30\textwidth}{!}{\includegraphics{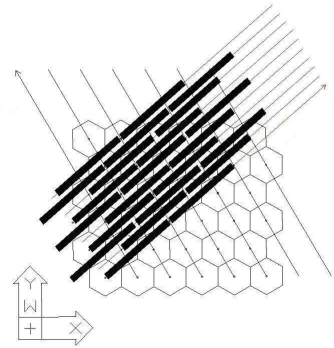}}
\caption{Sketch of the selected hexagonal configuration for the final spectra of
SPHERE IFS. The left-oriented axis is the reference on the IFU plane (filled with
hexagons representing a portion of the spaxels lattice), while the right-oriented
axis traces the dispersion direction and the black rectangles the spectra imaged
onto the detector plane. The spectra are 35 pixel long in the dispersion direction
and the separation to the nearest neighborhood is 5 pixel, both in the spectral
and in the spatial direction.}
\label{fg:Hexagonal-Conf}
\end{center}
\end{figure}

\section{Comparing different BIGRE and TIGER spaxel shapes and IFU lattice configurations}\label{sec:TIGERvsBIGRE}

In this Section we compare the slit functions generated through the TIGER and
BIGRE image propagation, computed for different spaxel shapes and lattice configurations
of the entire IFU. This comparison is made assuming common spaxel size and wavelength.
This analysis allows to derive the best lenslet-array optical concept and the optimum IFU
lattice configuration in the ideal diffraction limited case, i.e. when the object plane
of the lenslet-array is an un-resolved entrance pupil.

The diagnostic quantities exploited for this analysis are the amount of coherent and incoherent
intensities both measured onto the entrance slits plane of the spectrograph, before any chromatical
dispersion and re-imaging onto a suited detector plane. To this aim, it is important to stress
the meaning of coherent and incoherent signals and the one of their related cross-talk terms.
Coherent signal is the intensity term due to interference between adjacent spaxels measured
at any point of the entrance slits plane. Such a signal depends on the optical path difference
between adjacent spaxels only; in this sense spaxels can be compared to apertures of a
standard grating. The coherent cross-talk coefficient is the maximum amount of coherent
signal, see \S \ \ref{sec:CCT}. Incoherent signal is the stray intensity terms due to the image
propagation diffraction effects measured at any point of the entrance slits plane. Such a signal
depends on the distance between adjacent spectra projected onto this plane; in this sense this signal
depends on the final configuration of the spectra on to the detector plane.
The incoherent cross-talk coefficient is the maximum amount of incoherent signal,
see \S \ \ref{sec:ICT}.

The comparison between TIGER and BIGRE is performed for two distinct shapes of the single spaxel
(circular and square) and for two distinct IFU lattice configurations (hexagonal and square).
The combination of such different shapes and configurations allows to compare the TIGER and the BIGRE
concepts in term of coherent and incoherent signals for standard lenslet-array optical setups.
It is important to notice that these simulations consider as input a normalized signal
without amplitude and phase differences between adjacent spaxels. In this way the results
obtained are independent with respect to the actual speckle pattern beating the IFU.

\begin{figure*}
\centering
\begin{tabular}{c c c}
{\includegraphics[height=0.30\textwidth]{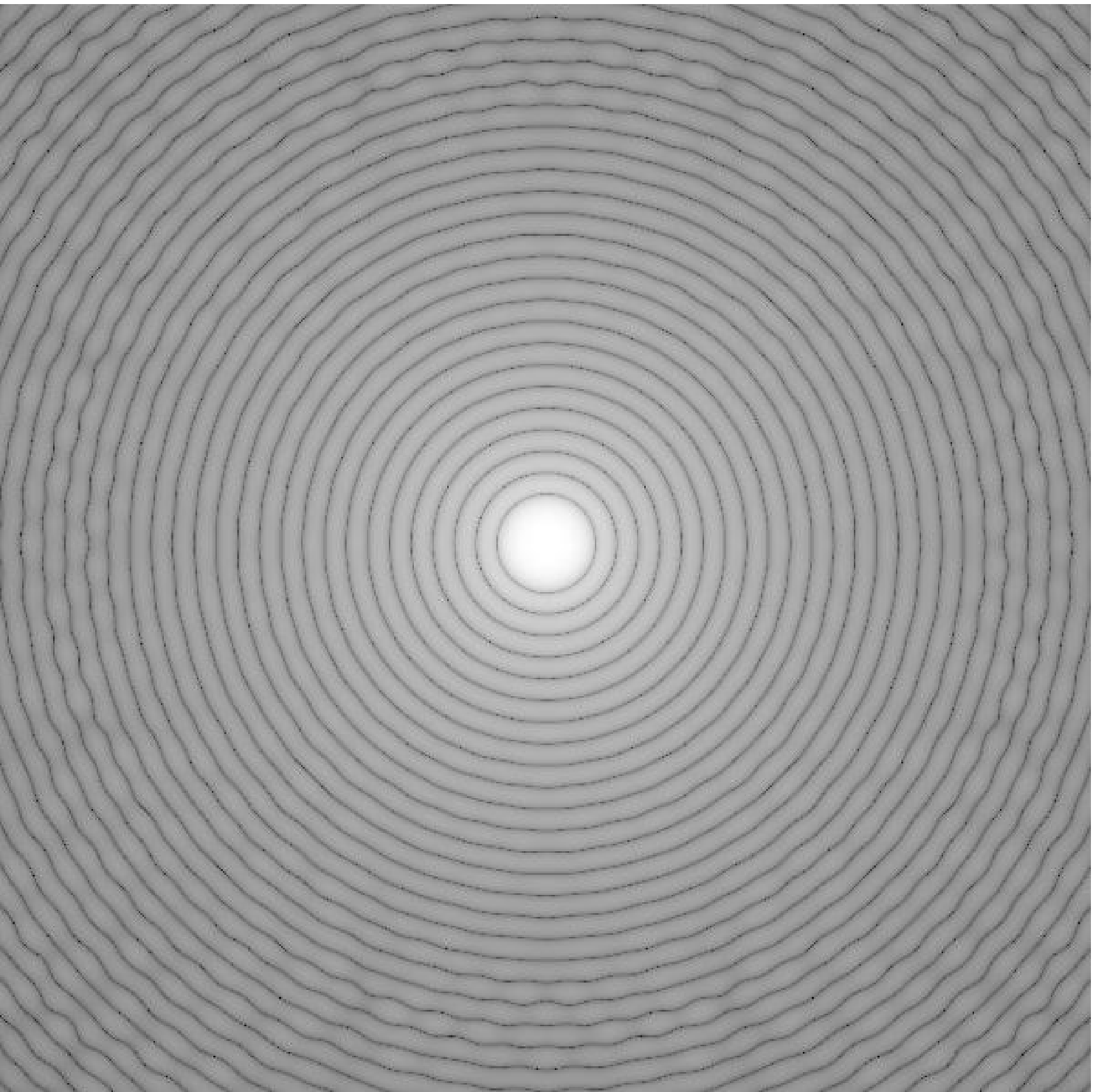}}    &
{\includegraphics[height=0.30\textwidth]{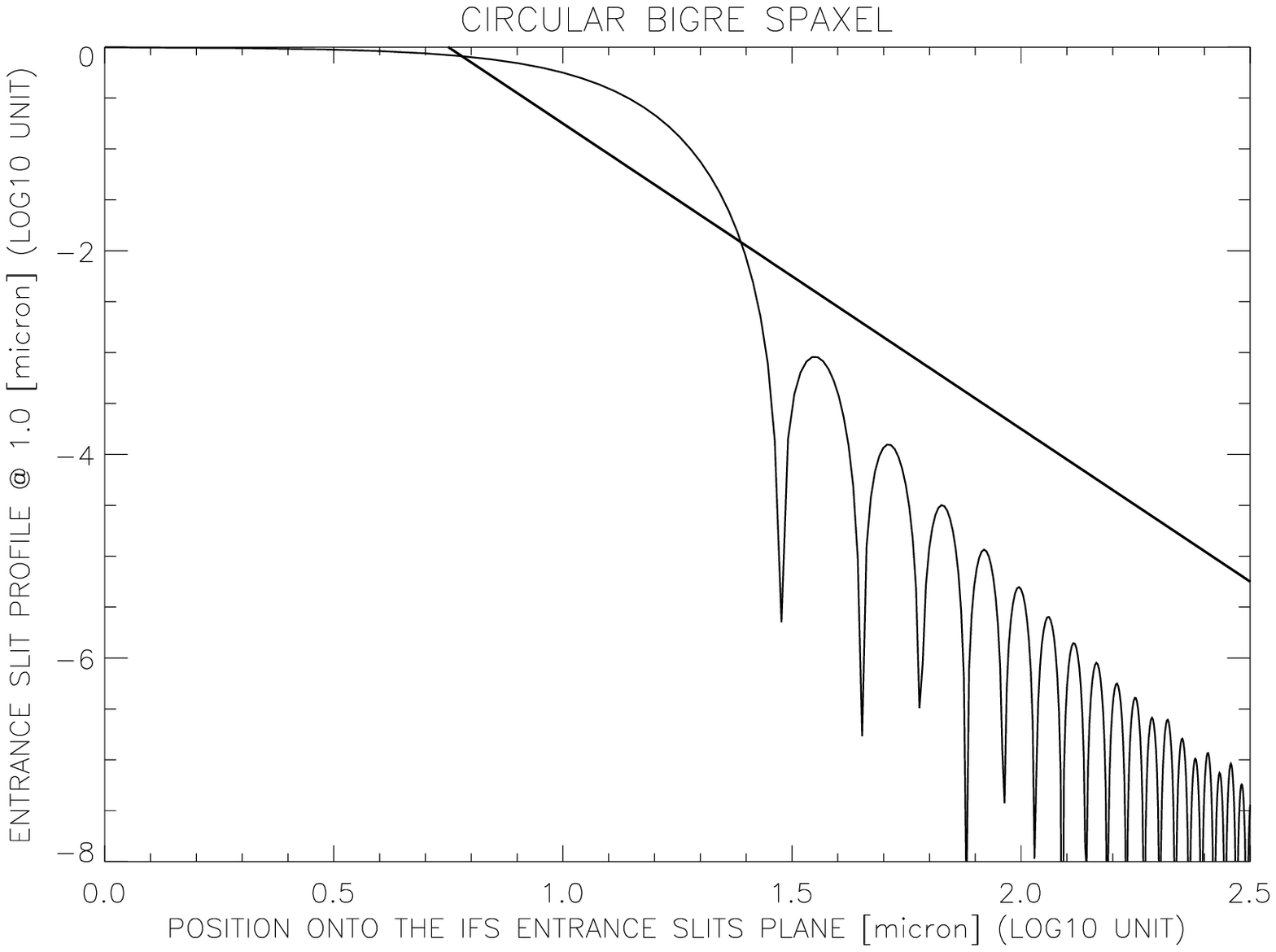}}    \\
{\includegraphics[height=0.30\textwidth]{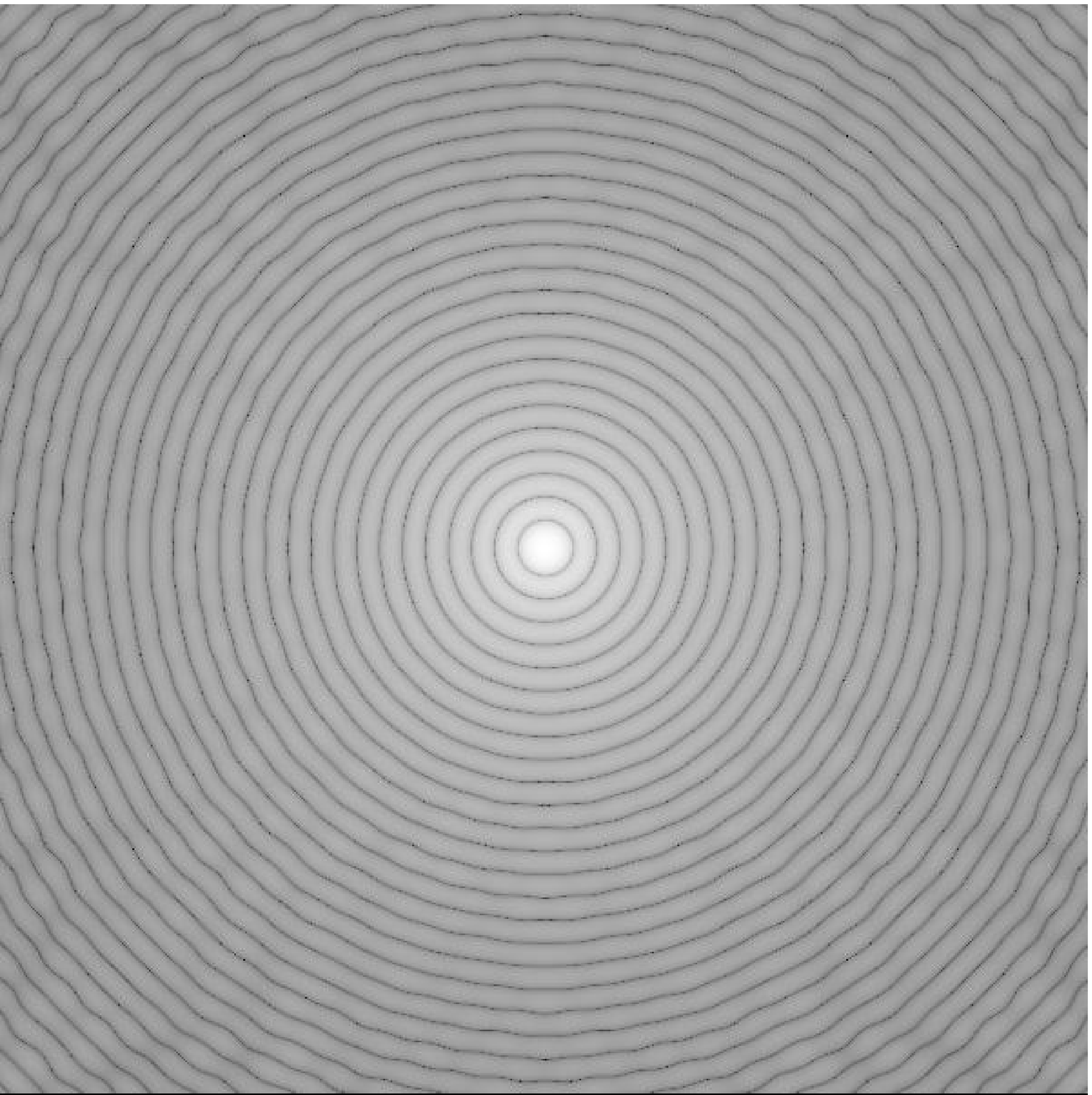}}    &
{\includegraphics[height=0.30\textwidth]{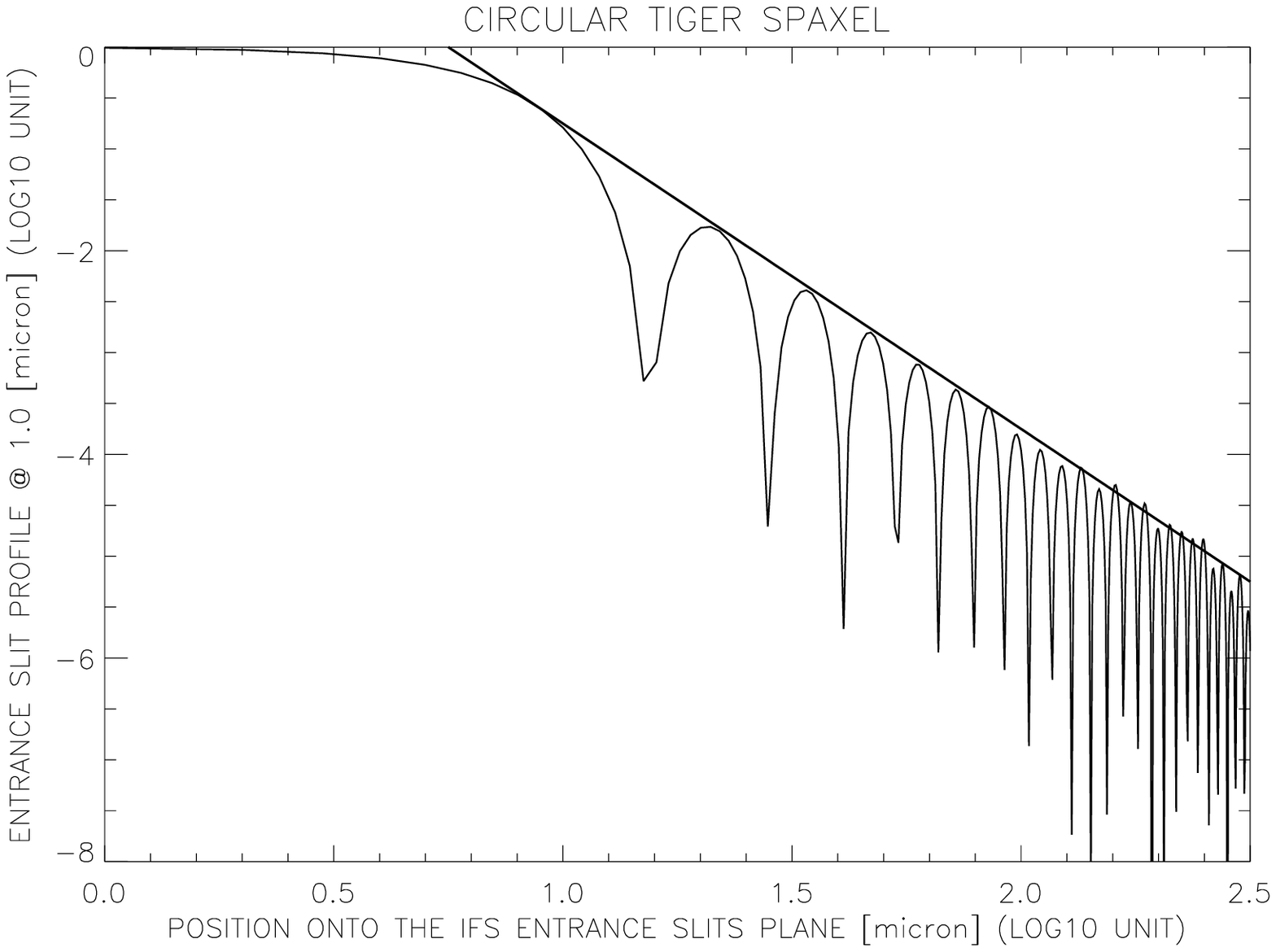}}    \\
{\includegraphics[height=0.30\textwidth]{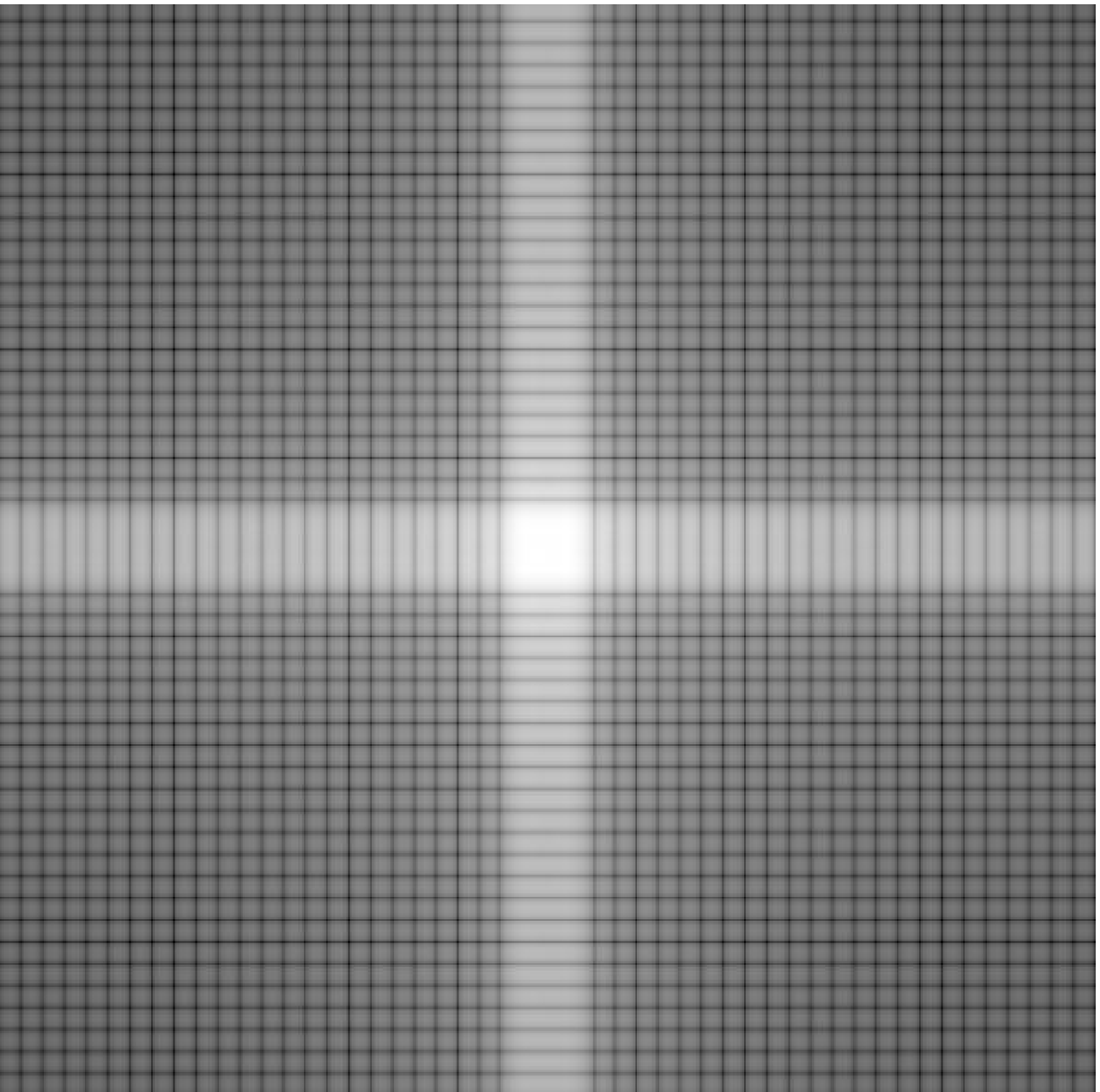}}    &
{\includegraphics[height=0.30\textwidth]{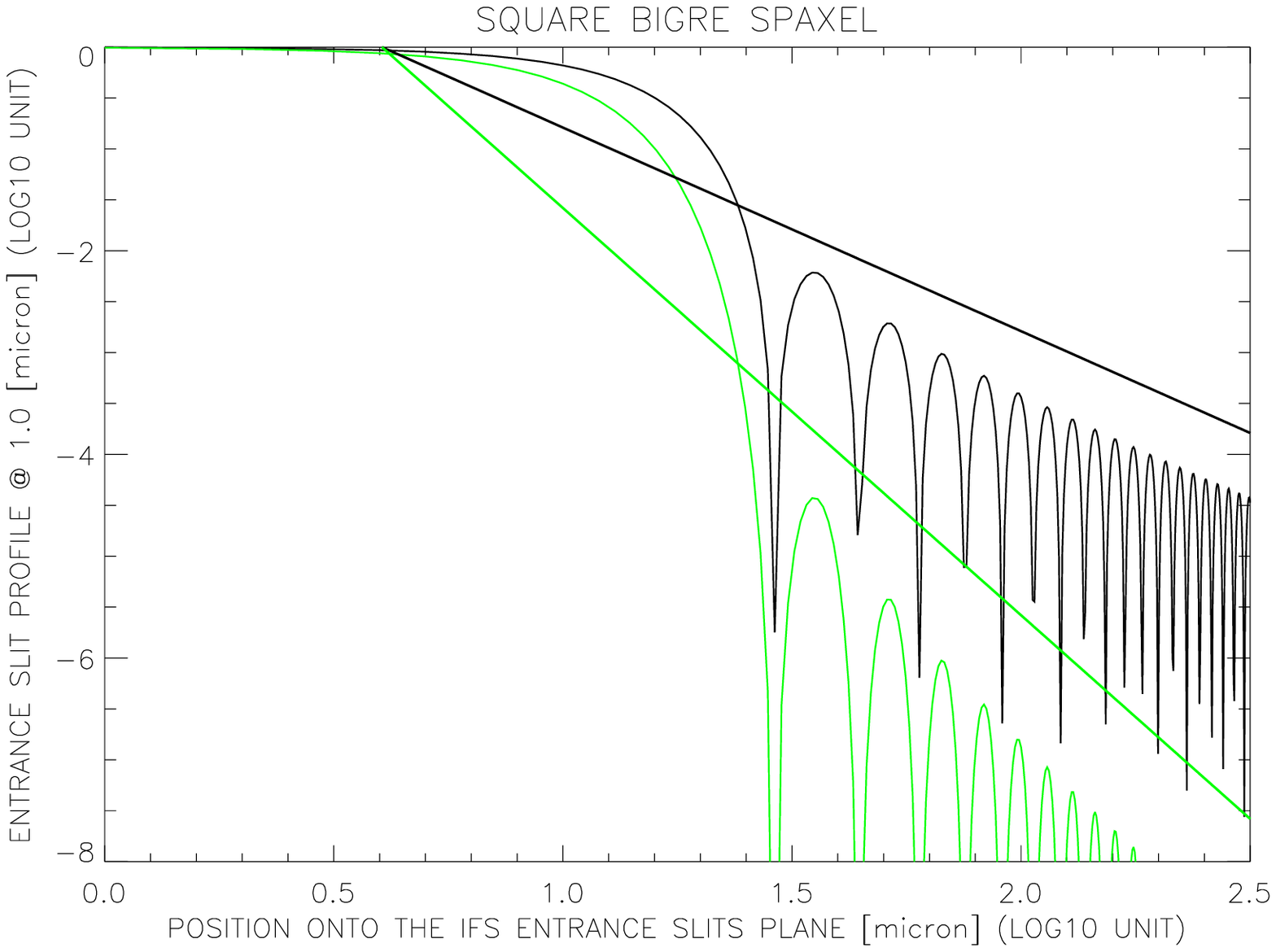}}    \\
{\includegraphics[height=0.30\textwidth]{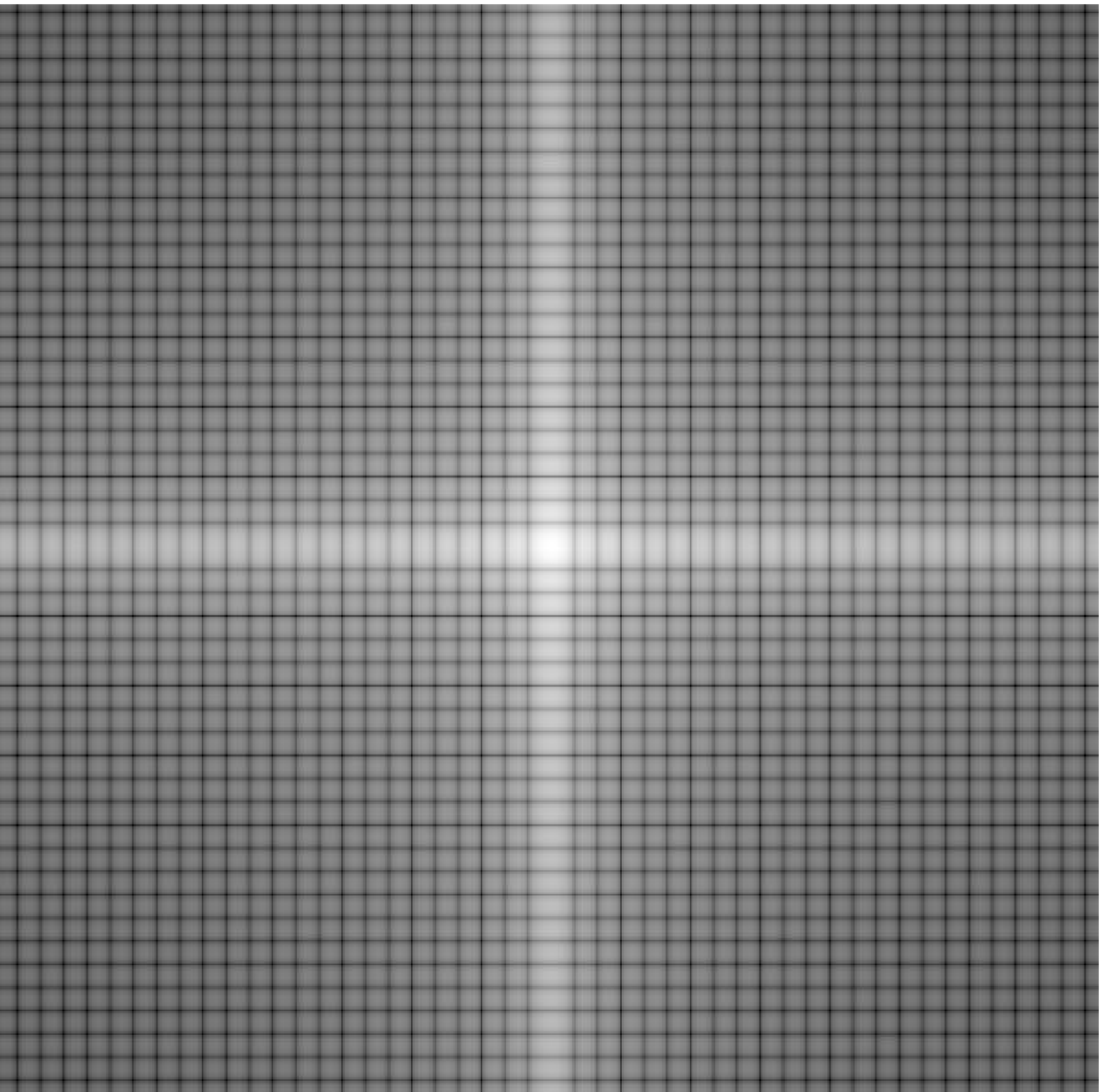}}    &
{\includegraphics[height=0.30\textwidth]{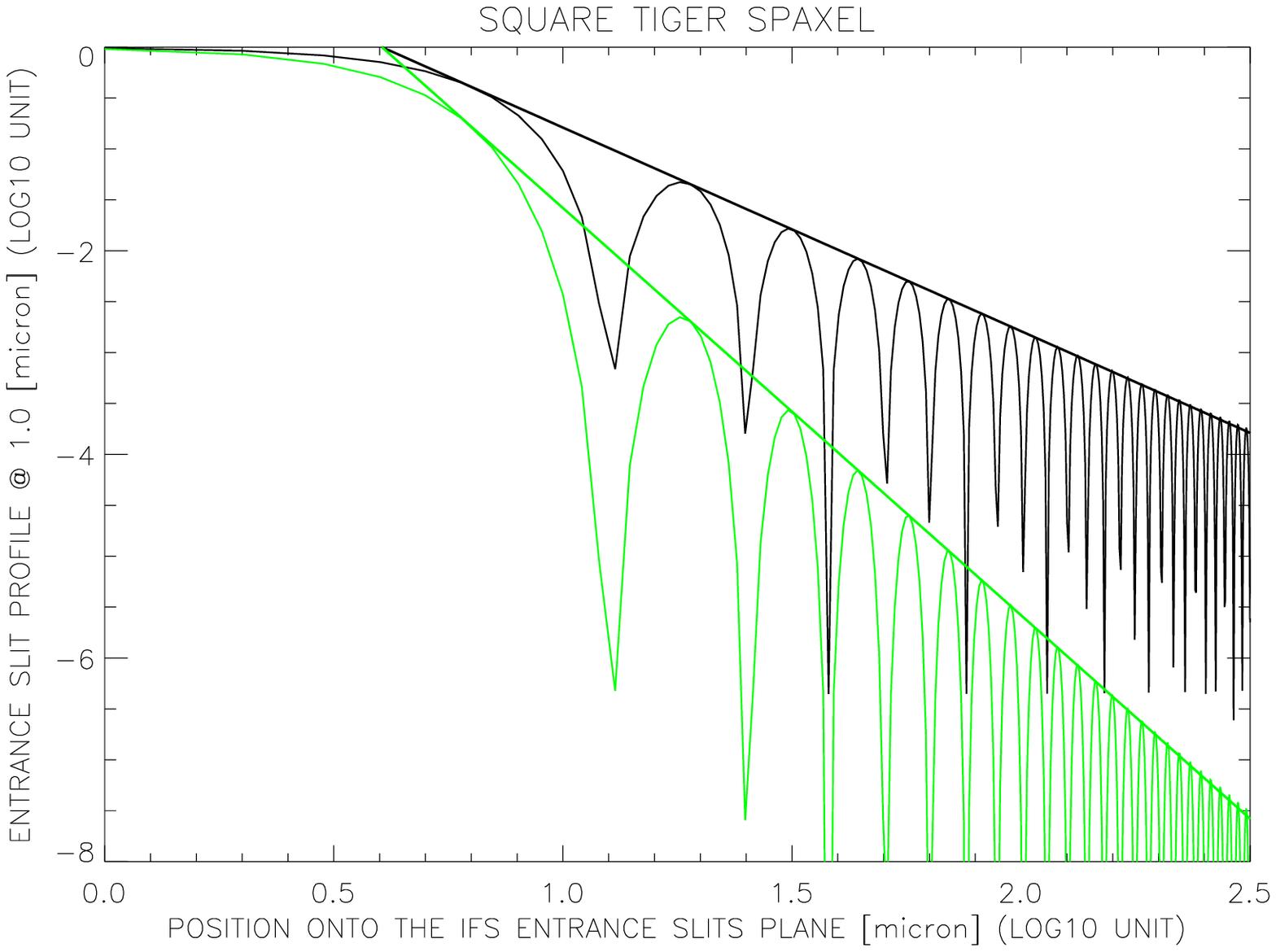}}    \\
\end{tabular}
\caption{Normalized BIGRE and TIGER slit functions comparison.
The adopted lenslet aperture is always $161.5 \ \mu m$ i.e. the one proper to the BIGRE solution
for SPHERE (see Table \ref{tab:bigre-ifs-parameters}); the adopted wavelength is always $1 \ \mu m$.
First column shows the images given by individual spaxels in a bi-logarithmic scale.
Second column shows the relative profiles (in the case of square spaxel shape diagonal
profiles are presented in green color). The power laws fitting the slit functions proper to
a cicular and a square TIGER-oriented spaxel are indicates for reference by solid lines. These simulations
consider Fraunhofer propagation only and assume as input signal of the lenslet-array an un-resolved entrance pupil.
Finally, results are independent to the detector pixel scale just because images refer to the spectrograph's entrance slits plane.}
\label{fg: TIGERvsBIGRE-1}
\end{figure*}

\begin{figure*}
\centering
\begin{tabular}{c c}
{\includegraphics[height=0.27\textwidth]{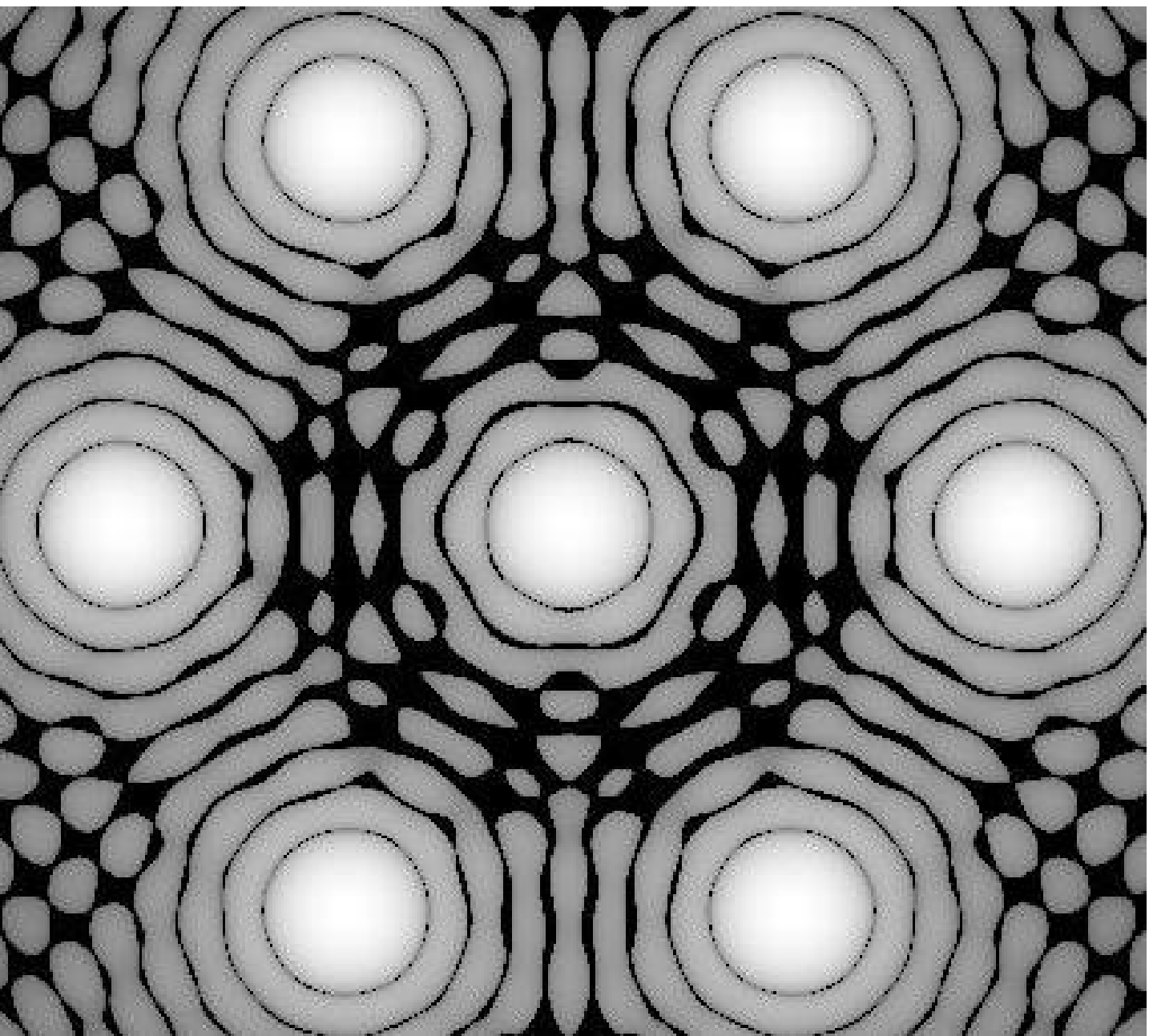}}    &
{\includegraphics[height=0.30\textwidth]{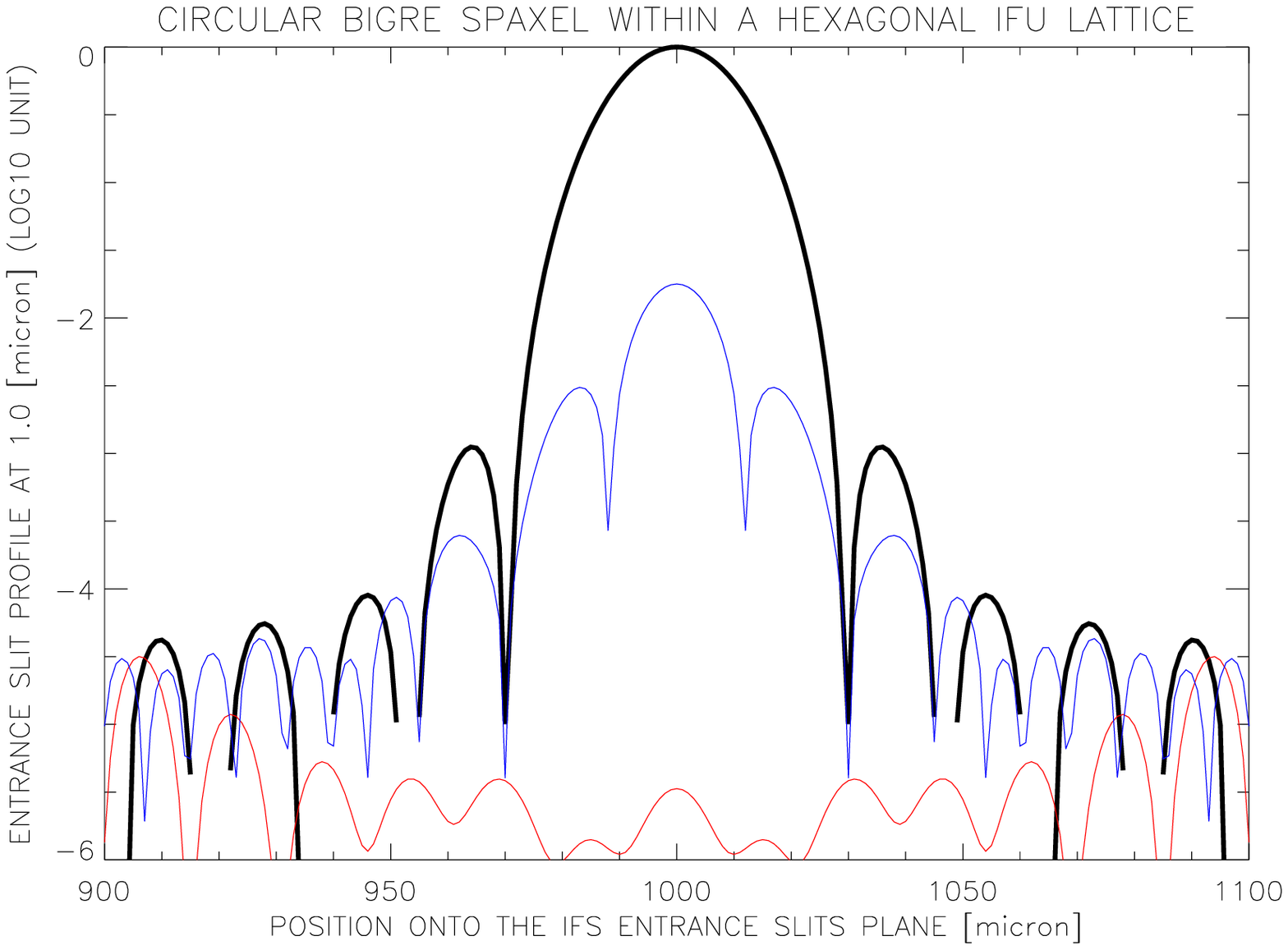}}    \\
{\includegraphics[height=0.30\textwidth]{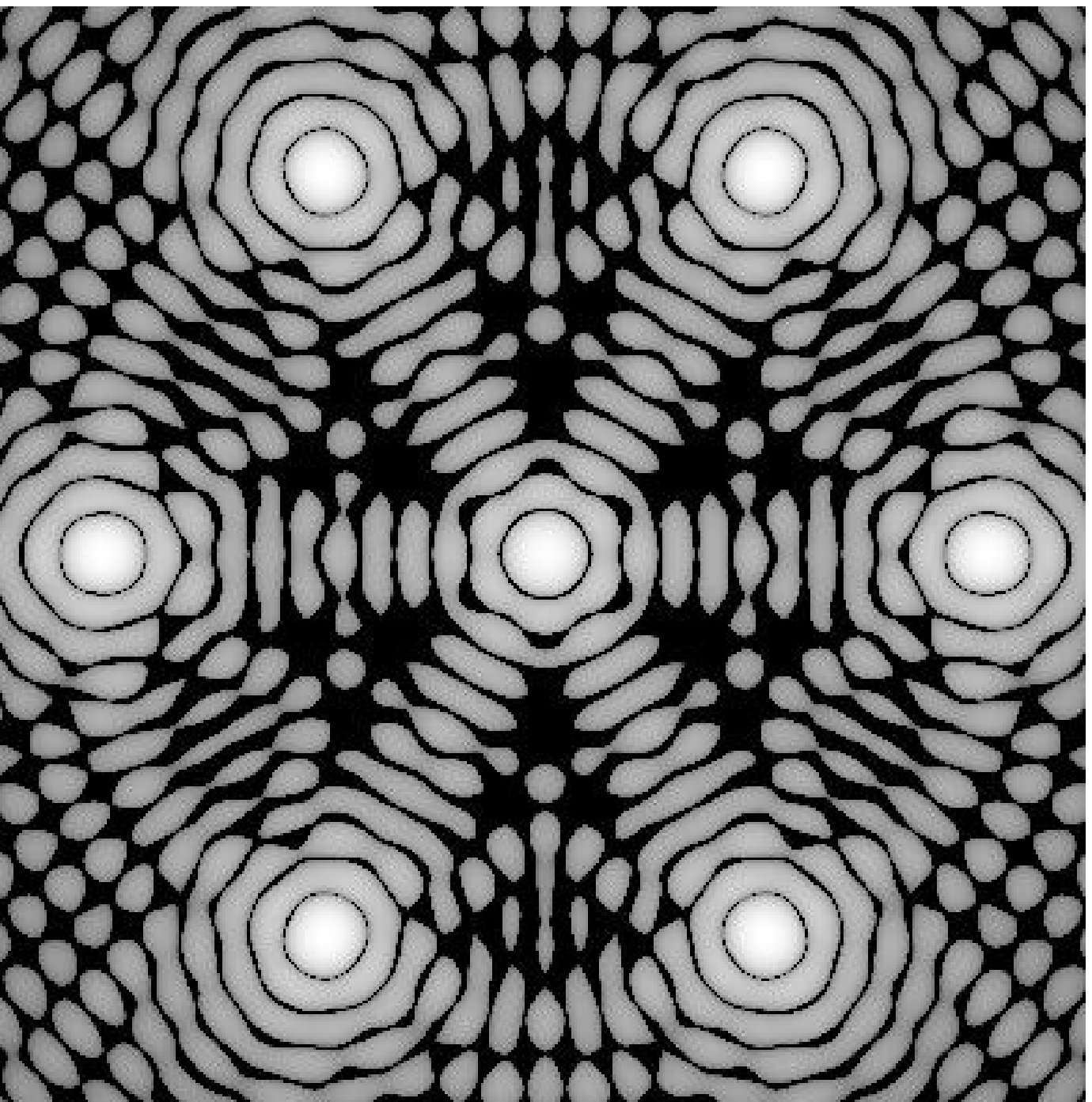}}    &
{\includegraphics[height=0.30\textwidth]{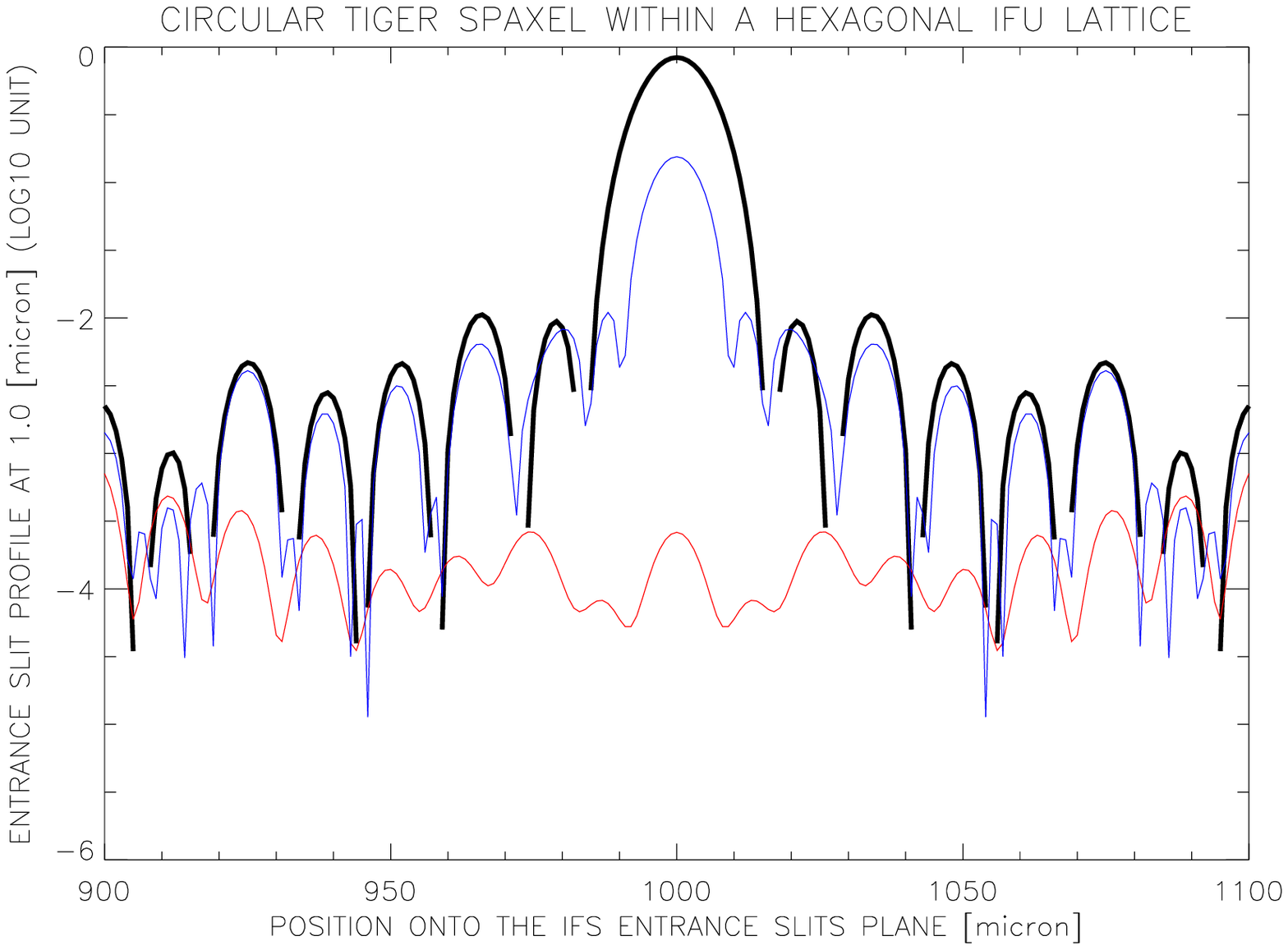}}    \\
{\includegraphics[height=0.30\textwidth]{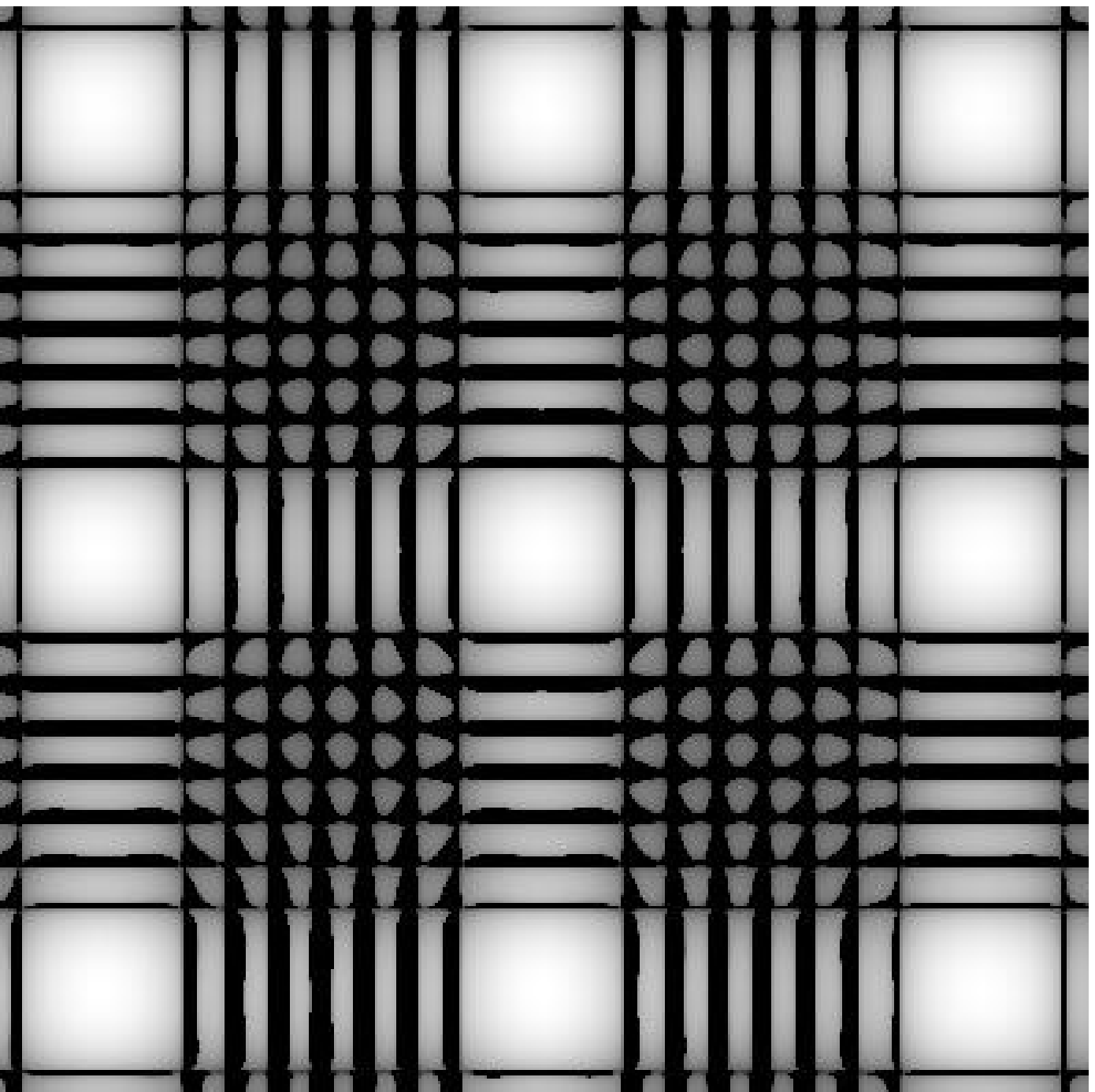}}    &
{\includegraphics[height=0.30\textwidth]{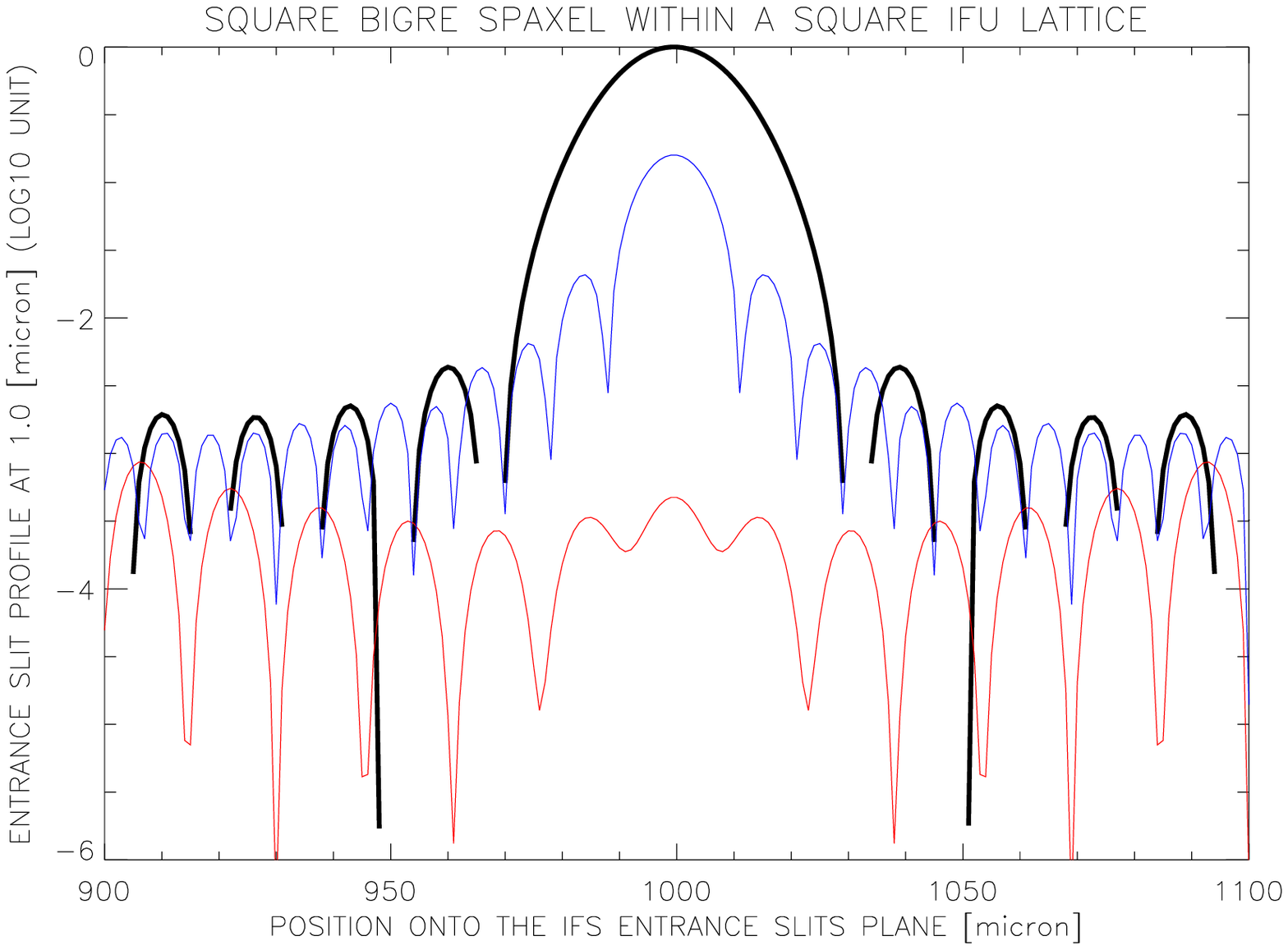}}    \\
{\includegraphics[height=0.30\textwidth]{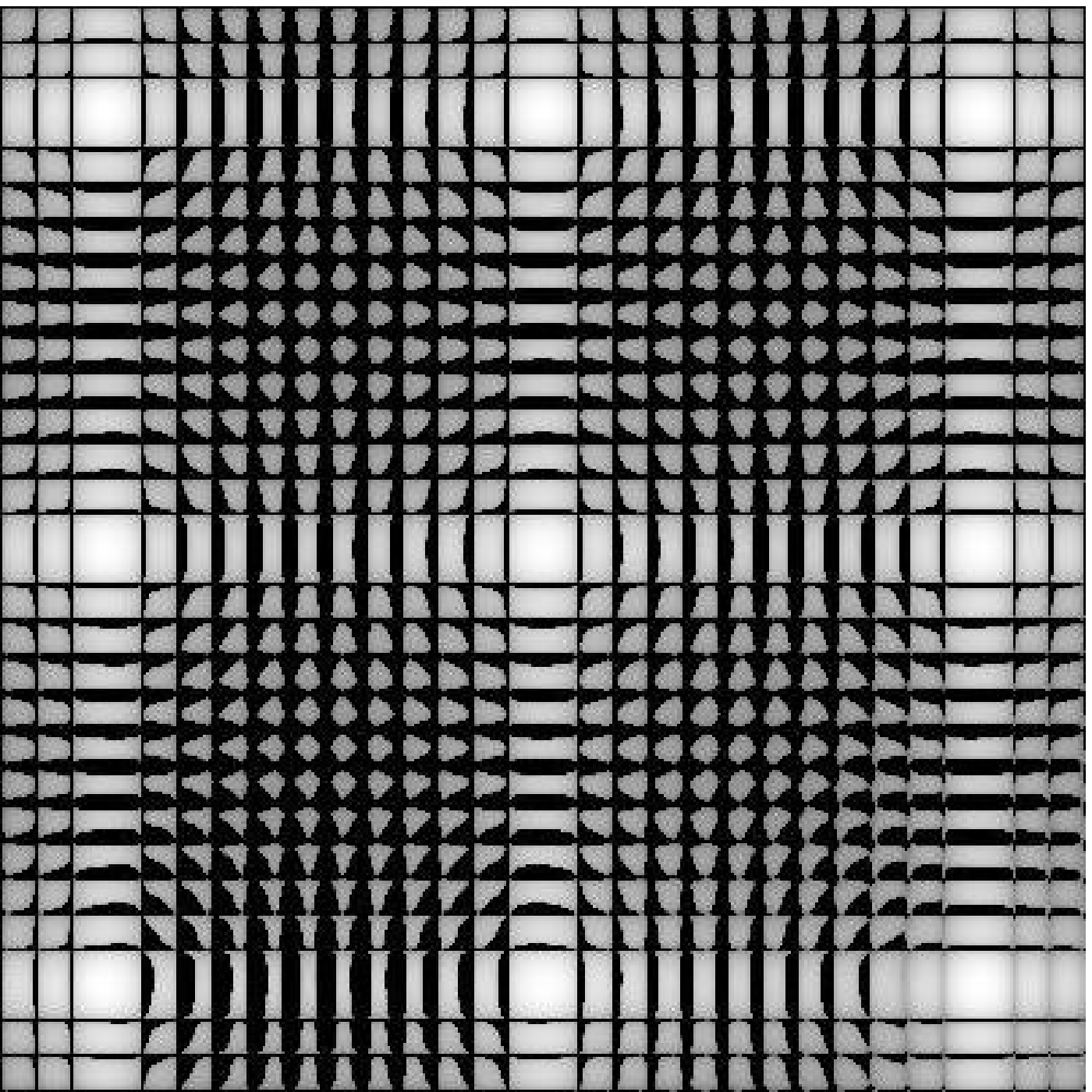}}    &
{\includegraphics[height=0.30\textwidth]{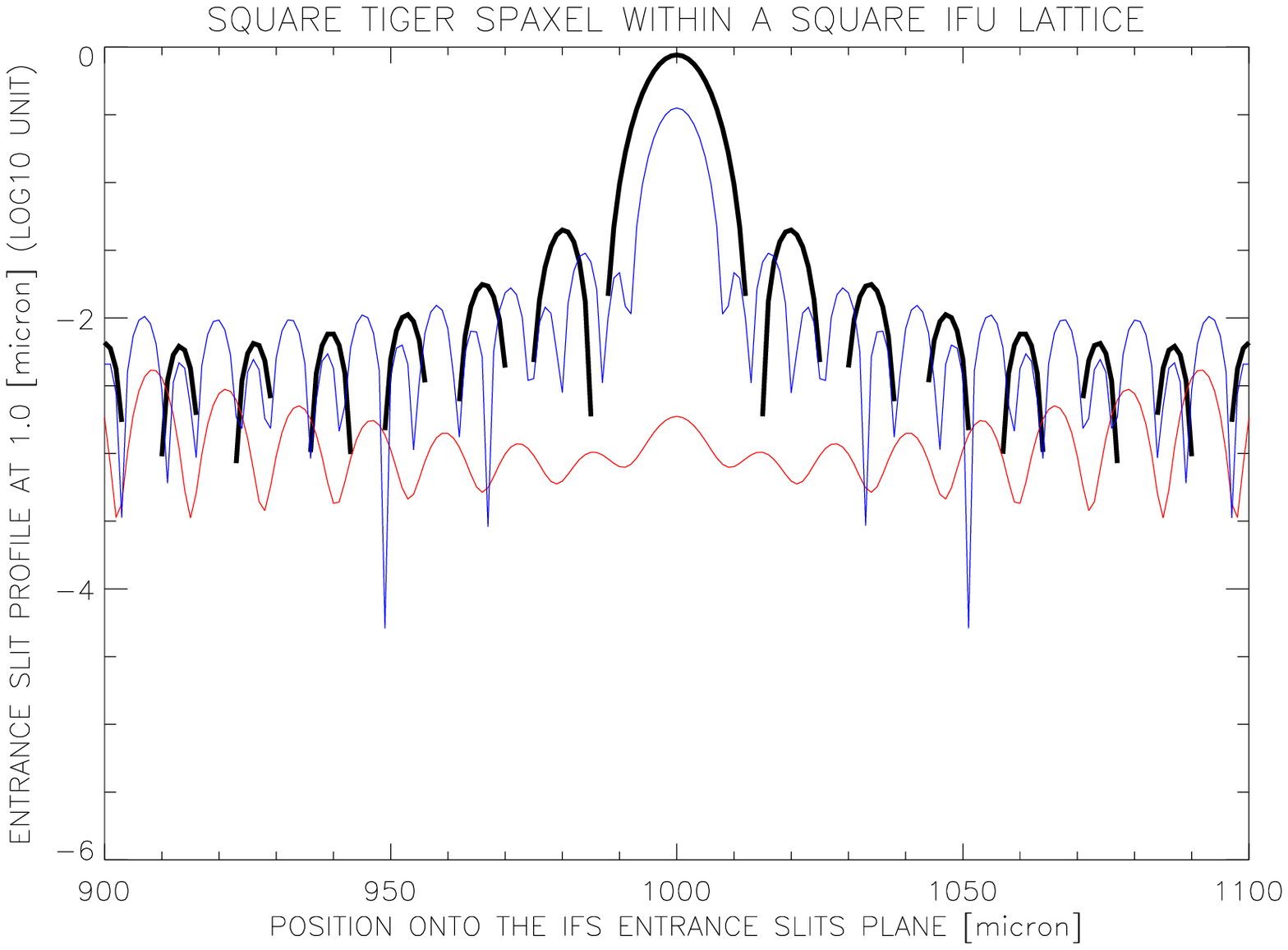}}    \\
\end{tabular}
\caption{Normalized BIGRE and TIGER slit functions comparison.
The adopted lenslet aperture is always $161.5 \ \mu m$ i.e. the one proper to the BIGRE solution
for SPHERE (see Table \ref{tab:bigre-ifs-parameters}); the adopted wavelength is always $1 \ \mu m$.
First column shows the images given by an individual spaxel together with its adjacent ones and for
different IFU lattice configurations. Second column shows the relative profiles: black color
indicates the slit function (the signal represents the detection expected if only the
central spaxel were illuminated); red color indicates the signal due to adjacent spaxels only;
blue color is the coherent signal arising from the interference of the signal
proper to the central spaxel with all the adjacent ones. These simulations consider Fraunhofer propagation only
and assume as input signal of the lenslet-array an un-resolved entrance pupil. Finally, results
are independent to the detector pixel scale just because images refer to the spectrograph's entrance slits plane.}
\label{fg: TIGERvsBIGRE-2}
\end{figure*}

As Figure \ref{fg: TIGERvsBIGRE-1} indicates, adopting a bi-logarithmic scale, the single BIGRE
slit gets an intensity profile steeper than the one proper to the single TIGER
slit both in the case of circular and square shapes. More in detail, the upper envelope to
the slit intensity profile proper to a circular TIGER-oriented spaxel
is a power law with index equal to $-3$, while the same quantity for a square
TIGER-oriented spaxel is a power law with index equal to $-2$ along the aperture side
and with index equal to $-4$ along its diagonal. At contrary, the upper envelope of the slit
intensity profile proper to a circular BIGRE-oriented spaxel is not a power law
(only its asymptotic tail is fitted quite well with a power law having index $\propto -4.5$);
the same quantity is not a power law in the case of a square BIGRE-oriented spaxel too (only its
asymptotic tail is fitted quite well with a power law with index $\propto -3$ in the direction
of the aperture side and index $\propto -6$ along its diagonal).

The result is that the BIGRE-oriented circular
aperture within a hexagonal lattice configuration allows a superior suppression
of coherent and incoherent signals, while the slits generated by a circular TIGER-oriented
aperture in a hexagonal lattice are similar --- in this context --- to the ones generated by a square
BIGRE-oriented aperture in a square lattice. Finally, the slits generated by a square TIGER-oriented
aperture in a square lattice are the worst in term of coherent and incoherent signals suppression,
see Figure \ref{fg: TIGERvsBIGRE-2}. Hence, the contribution of non-adjacent
spaxels can be neglected when evaluating the cross-talk signals in the case of a BIGRE spectrograph, just because
the power laws fitting --- in a bi-logarithmic plot --- the intensity distribution proper to the TIGER slit functions
do not fit at all the one proper to the BIGRE slit functions. At contrary, the intensity distribution proper to the
BIGRE slit functions can be only approximated with lower index power laws. Thus, what for
a TIGER lenslet-array represents an estimate only, for a BIGRE lenslet-array it gives
realistic measures of the signals due to the spectrograph's slit functions cross-talk.

\section{Conclusions}\label{sec:end}

By integral field spectroscopy it is possible to realize the S-SDI calibration technique
in the way proposed by \cite{BAl06}, and --- at least in a few cases --- to get the
spectrum of candidate extrasolar giant planets adopting suited spectral de-convolution
recipes, as the one proposed by \cite{T07}. However, these techniques can increase the contrast
performances only when several sampling conditions, both in the spatial and in the spectral
domain of the speckle field, are verified.

In this context, our effort has been to discuss in general terms
the critical sampling conditions needed to deal with a speckle field data cube
before applying on it the S-SDI calibration technique or any spectral de-convolution recipe.
To this purpose, we evaluated the impact of the cross-talk as function of various
parameters of a lenslet-based integral field spectrograph, especially in the case of trying to minimize
the number detector pixels (which is an issue in general for IFS) in the case of strong specifications,
as the ones requested for high-contrast imaging.
For this reason we conceived a new optical scheme --- we named BIGRE ---
and characterized it in the specific case of the IFS channel foreseen inside SPHERE, showing that a
BIGRE-oriented spectrograph is conceptually feasible by standard dioptric optical devices.
Once applied to the technical specifications of this instrument, a BIGRE
integral field unit is able to take into account the effects appearing if a lenslet-array
is used in diffraction-limited conditions. Specifically, we proved here that coherent
and incoherent cross-talk coefficients reach values deeper than for a TIGER IFU
when applied to the same optical frame. More in general, the comparison between the
BIGRE and the TIGER spaxel concept has been pursued
in terms of coherent and incoherent cross-talk suppression, adopting a common size for the single
aperture and a fixed monochromatic wavelength for the wavefront propagation. In the ideal case
of uniform illumination with un-resolved entrance pupil, the circular BIGRE spaxel
within an hexagonal IFU lattice configuration shows to be the optimal solution
among the ones we investigated.

\vspace{0.1truecm}
The authors thank Roberto Ragazzoni for the support he gave them in the development of this
subject, from the primeval CHEOPS project to SPHERE. Jacopo Antichi thanks personally Bernard Delabre
for a dedicated work session at ESO-Garching in April 2007, devoted to the final design optimization
of the BIGRE-oriented spectrograph to be mounted in SPHERE and Christophe V\'erinaud for his advising
during the completion of the manuscript. Jacopo Antichi is supported by LAOG through the European Seventh
Framework Programme INFRA-2007-2.2.1.28.

\end{document}